\documentclass[11pt,letterpaper]{article}
\usepackage[margin=1in]{geometry}

\usepackage{amssymb}
\usepackage{mathtools}
\usepackage{amsthm}
\usepackage{mathrsfs}
\usepackage{xcolor}
\usepackage[hidelinks]{hyperref}
\usepackage{changepage}
\usepackage{enumitem}
\usepackage{noindentafter}
\usepackage{etoolbox}
\usepackage{bookmark}
\usepackage{suffix}
\usepackage[toc,page]{appendix}

\usepackage{physics2}
\usephysicsmodule{braket}

\usepackage[
  style=alphabetic,
  url=false,
  isbn=false,
  backend=biber,
]{biblatex}
\AtEveryBibitem{\clearlist{language}}
\bibliography{refs.bib}

\setlist{itemsep=0pt}

\newenvironment{novelty*}{\begin{adjustwidth}{0cm}{0cm}\itshape}{\end{adjustwidth}\smallskip}
\NoIndentAfterEnv{novelty}
\NoIndentAfterEnv{novelty*}

\makeatletter
\newcommand{\etal}{et al\@ifnextchar.{}{.\@}}

\newcommand{\ie}{i.e.\@}
\newcommand{\etc}{etc\@ifnextchar.{}{.\@}}
\makeatother

\newcommand{\defn}[1]{\emph{#1}}

\newtheorem{theorem}{Theorem}[section]
\newtheorem{lemma}[theorem]{Lemma}
\newtheorem{corollary}[theorem]{Corollary}
\newtheorem{proposition}[theorem]{Proposition}
\newtheorem{claim}[theorem]{Claim}

\theoremstyle{definition}
\newtheorem{definition}[theorem]{Definition}

\theoremstyle{remark}
\newtheorem*{remark}{Remark}

\patchcmd{\thmhead}{(#3)}{#3}{}{}

\makeatletter
\newenvironment{innerproof}[1][\proofname]
  {\par\normalfont \topsep6\p@ \@plus6\p@\relax
  \trivlist
  \item[\hskip\labelsep\itshape#1\@addpunct{.}]\ignorespaces}
  {\endtrivlist\@endpefalse}
\makeatother

\renewcommand{\phi}{\varphi}
\renewcommand{\epsilon}{\varepsilon}

\newcommand{\Z}{\mathbb{Z}}
\newcommand{\R}{\mathbb{R}}
\newcommand{\C}{\mathbb{C}}

\newcommand{\caseif}{\text{if}~}

\newcommand{\abs}[1]{\left|#1\right|}
\WithSuffix\newcommand\abs*[1]{\lvert#1\rvert} %

\newcommand{\poly}{\mathit{poly}}

\DeclarePairedDelimiter{\ceil}{\lceil}{\rceil}

\newcommand{\restr}[2]{{
  \left.\kern-\nulldelimiterspace
  #1
  \littletaller
  \right|_{#2}
  }}
\newcommand{\littletaller}{\mathchoice{\vphantom{\big|}}{}{}{}}

\newcommand{\logic}[1]{\mathscr{#1}}

\newcommand{\FPC}{\mathsf{FPC}}

\newcommand{\str}[1]{\mathcal{#1}} %
\newcommand{\strb}[1]{\langle#1\rangle}
\newcommand{\fin}{\mathrm{fin}}
\newcommand{\taugraph}{\tau_{\mathrm{graph}}}
\newcommand{\prop}[1]{\mathscr{#1}} %
\newcommand{\PCFI}{\prop{P}_{\mathrm{CFI}}}
\newcommand{\PAltCFI}{\prop{P}_{\mathrm{CFI}}^{\mathrm{Alt}}}

\newcommand{\Sym}[1]{\mathrm{Sym}_{#1}}
\newcommand{\Alt}[1]{\mathrm{Alt}_{#1}}
\DeclareMathOperator{\Stab}{Stab}

\DeclareMathOperator{\Orb}{Orb}
\DeclareMathOperator{\Aut}{Aut}
\DeclareMathOperator{\In}{In}

\newcommand{\Hilb}{\mathcal{H}}
\newcommand{\tprod}{\otimes}

\newcommand{\iden}{\mathbb{I}}
\DeclareMathOperator{\Tr}{Tr}
\newcommand{\TrTo}[1]{\Tr_{\overline{#1}}} %
\DeclareMathOperator{\DTr}{DTr} %
\newcommand{\DTrTo}[1]{\DTr_{#1}}
\newcommand{\Dens}{\mathrm{D}} %
\newcommand{\Unit}{\mathrm{U}} %
\newcommand{\ew}{\eta} %

\makeatletter
\let\oldket\ket
\def\ket{\@ifstar{\oldket}{\oldket*}}
\let\oldbra\bra
\def\bra{\@ifstar{\oldbra}{\oldbra*}}
\let\oldbraket\braket
\def\braket{\@ifstar{\oldbraket}{\oldbraket*}}
\let\oldketbra\ketbra
\def\ketbra{\@ifstar{\oldketbra}{\oldketbra*}}
\makeatother

\newcommand{\Obs}{\mathcal{O}} %
\newcommand{\ClasObs}{\mathcal{O}_{\mathrm{clas}}} %
\newcommand{\Peb}{\mathcal{P}} %
\newcommand{\inter}{{\mathchoice{\textstyle}{}{}{}\bigcap}}

\newcommand{\gateset}{\mathbb{G}}

\newcommand{\CNOT}{\mathrm{CNOT}}

\newcommand{\out}{\mathrm{out}}

\newcommand{\layer}{\mathcal{L}}
\DeclareMathOperator{\minsupp}{\mathbf{min-supp}}
\DeclareMathOperator{\maxsupp}{\mathbf{max-supp}}
\DeclareMathOperator{\maxorb}{\mathbf{max-orbit}}
\DeclareMathOperator{\supp}{\mathit{sp}} %

\newcommand{\Wd}[1]{W^{\underline{#1}}} %
\newcommand{\concat}{\oplus}

\newcommand{\footremember}[2]{%
    \footnote{#2}
    \newcounter{#1}
    \setcounter{#1}{\value{footnote}}%
}
\newcommand{\footrecall}[1]{%
    \footnotemark[\value{#1}]%
}

\numberwithin{equation}{section}
\allowdisplaybreaks
\hypersetup{
  pdftitle={Entanglement Lower Bounds for Symmetric Quantum Circuits},
  pdfauthor={Davi Castro-Silva and Ryan Dancy and Sergii Strelchuk}
}
\title{Entanglement Lower Bounds for Symmetric Quantum Circuits}
\author{Davi Castro-Silva\footnote{University of Warwick} \and Ryan Dancy\footremember{oxford}{University of Oxford} \and Sergii Strelchuk\footrecall{oxford}}
\date{}
\begin{document}

\hypersetup{pageanchor=false}
\pagenumbering{gobble}
\maketitle
\thispagestyle{empty}
\begin{center}
\textbf{Abstract}
\end{center}

Symmetric circuits require every symmetry of the input to extend to an automorphism of the circuit. In the classical setting, their connection to descriptive complexity yields strong unconditional lower bounds. We extend this approach to the quantum model of symmetric computation defined in~\cite{CastroSilva2026}, proving a tradeoff between circuit size and entanglement width (the smallest $b$ such that, on every computational-basis input, the state after each layer factors into states on at most $b$ qubits each).

Our main result identifies a graph property decidable in classical polynomial time that no polynomial-size family of symmetric quantum circuits on $n$-vertex graphs with entanglement width $o(n/\log n)$ can compute.
The bound holds for circuits with arbitrary one-qubit and two-qubit gates and unrestricted depth, and remains valid under the weaker notion of symmetry with respect to even permutations of vertices. Our proof is based on a new quantum pebble game based on the partial trace. The resulting indistinguishability theorem transfers classical pebble-game lower bounds to quantum circuits with restricted entanglement.

By contrast, without entanglement restrictions, we show that every graph property in BQP can be computed by a uniform family of polynomial-size symmetric quantum circuits. Restricting to circuits with no entanglement at all, we show that polynomial-size symmetric quantum circuits with entanglement width~1 can compute every graph property definable in fixed-point logic with counting. Finally, we give an efficient symmetric implementation of QAOA for MAX 3XOR, providing a starting point for studying entanglement lower bounds in quantum optimization algorithms.

\clearpage

\pagenumbering{arabic}
\hypersetup{pageanchor=true}
\setcounter{page}{1}

\section{Introduction}\label{sec:introduction}

Graph properties such as 3-colorability and Hamiltonicity give rise to fundamental computational problems, with numerous applications in both theory and practice. Although these properties are, by definition, invariant under relabelling of vertices, algorithms that decide them need not respect these symmetries: they may single out an arbitrary vertex or use the vertex labels to impose an ordering. This mismatch between the levels of abstraction at which computational problems are formulated and at which algorithms are designed and analyzed was a central motivation for the study of \emph{choiceless polynomial time} by Blass, Gurevich, and Shelah~\cite{Blass1999}, and is closely related to the longstanding problem of capturing polynomial-time computation in descriptive complexity theory.

This line of research has led to the systematic study of \emph{symmetric circuits}, a model of computation in which the symmetries inherent to a problem are required to extend to the structure of the computation itself. Symmetric circuits thus provide a natural framework for investigating when symmetry-breaking, a form of algorithmic choice, is necessary for efficient computation.

Symmetric circuits were shown to admit exact characterizations in terms of mathematical logic, establishing a deep connection with descriptive complexity theory. In particular, Anderson and Dawar~\cite{Anderson2017} established a correspondence between uniform families of polynomial-size symmetric circuits (with threshold gates) and definability in fixed-point logic with counting ($\FPC$). This connection has made techniques from finite model theory available for studying circuit complexity, most notably \emph{pebble games} between Spoiler and Duplicator. Building on these ideas, Dawar~\cite{Dawar2015} and Dawar and Wilsenach~\cite{Dawar2024a,Dawar2024} developed methods for deriving symmetric circuit lower bounds from winning strategies for Duplicator in pebble games.

With a view to extending this programme to quantum computation, Castro-Silva, Gur, and Strelchuk~\cite{CastroSilva2026} introduced the notion of \emph{symmetric quantum circuits}. They showed how such circuits can efficiently implement fundamental quantum algorithmic techniques and, moreover, demonstrated exponential quantum advantage in the symmetric setting. In this paper, we turn to the complementary question: can the symmetry restriction itself be exploited to prove \emph{lower bounds} for quantum circuits? Developing such a technique would be particularly valuable given the scarcity of methods for establishing quantum circuit lower bounds.

Here, we introduce a quantum-inspired generalization of pebble games tailored to the study of symmetric quantum circuits. Our notion of \emph{quantum pebble games} takes a group-theoretic perspective, interpreting pebbles as restrictions on the symmetries available to Duplicator. The classical notion of partial isomorphism is replaced by a quantum analogue formulated in terms of reduced density matrices, recovering the classical notion for computational-basis states. By recasting pebble games in terms of symmetries and the local structure of quantum states, our framework offers a new perspective that we believe may be of independent interest.

Our main technical contribution is to turn this framework into a method for proving superpolynomial lower bounds on symmetric quantum circuits with restricted entanglement. In particular, we establish such lower bounds for an explicit graph property that is decidable in classical polynomial time: no family of polynomial-size symmetric quantum circuits with entanglement width $o(n/\log n)$ can compute this property. Thus, even for classically tractable problems, symmetry together with restrictions on entanglement can impose fundamental limitations on the power of quantum computation.

To complement this result, we show that bounding some quantum resource is needed for a result of this type to hold: without such restrictions, polynomial-size symmetric quantum circuits capture the full power of BQP on graph problems (and on more general relational structures).
Even in the total absence of entanglement, however, we prove that polynomial-size symmetric quantum circuits using only two-qubit gates can still compute all $\FPC$-definable properties.
This sharply contrasts with the classical setting, where polynomial-size symmetric circuits with Boolean gates cannot even compute the parity of the number of edges \cite{Dawar2015a}.

Finally, we give a direct symmetric implementation of QAOA for MAX 3XOR, together with a test of whether an assignment satisfies a prescribed number of equations. The resulting symmetric circuit accepts with the same probability that QAOA finds an assignment satisfying at least the prescribed number of equations.

\subsection{Main results}
Our results will be described in terms of the graph input model. 
A graph on $[n]$ is encoded by its $n^2$ adjacency bits in the computational basis, with one input qubit for each ordered pair of vertices. Additional ancilla qubits begin in $\ket{0}$. A circuit applies layers of pairwise commuting gates and measures a distinguished output qubit. Its size $s(C)$ is the total number of gates. Computing a property means producing the correct Boolean answer with probability at least $2/3$ on every input.

Let $\Sym{n}$ be the group of permutations of $[n]$. Graph properties are invariant under the action of $\Sym{n}$ on its vertices.
Quantum circuits are said to be \emph{symmetric} if they are likewise invariant under these symmetries:
we require every $\sigma\in\Sym{n}$ to extend from the input qubits, by $(u,v)\mapsto(\sigma(u),\sigma(v))$, to a permutation of the circuit wires that preserves each layer and fixes the output wire.
(See Section~\ref{sec:symmetric-circuits} for a precise definition.)

The entanglement width\footnote{Also known as \emph{entanglement depth}, e.g. \cite[Def.~18.7]{Bertlemann2023}.} of a state $\ket{\psi}$ is the smallest number $b$ such that $\ket{\psi}$ can be written as a tensor product of states on disjoint sets of at most $b$ qubits.
For example, a product state has width $1$, while a tensor product of Bell pairs has width $2$ regardless of the number of pairs.
The \defn{entanglement width} $\eta(C)$ of a quantum circuit $C$ is the smallest integer such that, on every computational-basis input and after each circuit layer, the global state of all wires has entanglement width at most $\eta(C)$;
the partition of the state into blocks is allowed to depend on both the input and the layer.
Formal definitions appear in Section~\ref{sec:model}.

The hard property $\PAltCFI$ is a fixed polynomial-time graph property separating two families obtained from the construction of Cai, F\"urer, and Immerman~\cite{Cai1992}. Their members are nonisomorphic but remain indistinguishable in an appropriate pebble game with linearly many pebbles. Section~\ref{sec:cfi} describes these families and the separating property.

We note that the property $\PAltCFI$ admits a classical polynomial-time algorithm. Our main result identifies a restriction to computing it symmetrically, even in the quantum setting: polynomial-size circuits must generate near-linear entanglement (on the number of vertices) in order to compute it.

\begin{theorem}\label{thm:main}
Let $f:\Z_+\to\Z_+$ be any increasing function with $f(n)=\omega(\log n)$. There is no family of symmetric quantum circuits $(C_n)_n$ using one-qubit and two-qubit gates computing $\PAltCFI$ which has size $s(C_n) = 2^{o(f(n))}$ and entanglement width $\eta(C_n) = O(n/f(n))$.
\end{theorem}

In this theorem, the function $f$ relates the allowed entangled block size to the excluded circuit size. If the blocks have size $O(n/f(n))$, a family computing the property cannot have gate count $2^{o(f(n))}$. The condition $f(n)=\omega(\log n)$ ensures that this excluded range includes every polynomial gate count.
We note that the exact result we prove is stronger, as it also holds for circuits which are only symmetric with respect to the alternating group $\Alt{n}$ of even permutations;
see Theorem~\ref{thm:mainAlt}.
As previously mentioned, the theorem is proven via a more general framework of Spoiler--Duplicator pebble games, which we describe in Section~\ref{subsec:pebble-games}. The following immediate consequences give the polynomial-size case and the case of a power-law bound on the block size.
\begin{corollary}\label{cor:poly}
There is no polynomial-size family of symmetric quantum circuits $(C_n)_n$ using one-qubit and two-qubit gates that computes $\PAltCFI$ and has entanglement width $\eta(C_n) = o(n/\log n)$.
\end{corollary}

\begin{corollary}\label{cor:subexp}
Let $0<\gamma\leq 1$. There is no family of symmetric quantum circuits $(C_n)_n$ using one-qubit and two-qubit gates computing $\PAltCFI$ which has size $s(C_n) = 2^{o(n^\gamma)}$ and entanglement width $\eta(C_n) = O(n^{1-\gamma})$.
\end{corollary}

Our next result shows that restricting some quantum resource is essential for obtaining the lower bounds above: unrestricted polynomial-time symmetric computation attains the full power of BQP.

\begin{theorem}
\label{thm:symbqp-simulation}
Let $\prop{P}$ be a graph property.
Suppose that a polynomial-time uniform family of quantum circuits computes $\prop{P}$ using $q(n)$ one-qubit and two-qubit gates on inputs of order $n$.
Then, $\prop{P}$ can also be computed by a polynomial-time uniform family of \emph{symmetric} quantum circuits using $O(q(n)+ n^4 \log^2(n))$ one-qubit and two-qubit gates.
\end{theorem}

Furthermore, restricting entanglement width does not trivialize our model of computation. Fixed-point logic with counting ($\FPC$) extends first-order logic with inductive definitions and counting, without assuming an ordering of the input vertices. It can express connectivity and the existence of a perfect matching, and captures all polynomial-time graph properties on planar graphs~\cite{Dawar2015a}. Our next result shows that every $\FPC$-definable graph property can be computed by a polynomial-time uniform family of polynomial-size symmetric quantum circuits of entanglement width $1$, using only one-qubit and two-qubit gates.
\begin{theorem}
\label{thm:ew-1-computes-fpc}
Let $\prop{P}$ be an $\FPC$-definable graph property.
Then $\prop{P}$ can be computed by a polynomial-time uniform family of symmetric quantum circuits with entanglement width 1 using only one-qubit and two-qubit gates.
\end{theorem}

Finally, we connect our framework to quantum optimization by giving an efficient symmetric implementation of the Quantum Approximate Optimization Algorithm which we apply to MAX 3XOR. An instance of this problem is given by $J=(A,d)$, and consists of $m$ equations $Ay=d$ over $\mathbb F_2$ with $n$ Boolean variables, where each row of $A$ contains three nonzero entries. The aim is to find an assignment satisfying as many equations as possible. QAOA begins with the uniform superposition of all assignments and at each round applies phases determined by the number of satisfied equations, followed by the application of $e^{-i\beta_r X}$ on every variable qubit. The parameters $\gamma_r$ and $\beta_r$ control these two operations in round $r$, respectively.

Permuting the equations permutes the rows of $A$ and the
corresponding entries of $d$, permuting the variables permutes
the columns of $A$. We call a circuit symmetric for this problem if every such independent permutation of the equations and variables extends to a permutation of all circuit wires that preserves each layer and fixes the output wire.

\begin{theorem}[(Symmetric implementation of QAOA)]
\label{thm:qaoa-compilation}
For positive integers $m,n, p\in\mathbb{N}$, a threshold $t\in\{0,\ldots,m\}$, and angles $\boldsymbol{\gamma}=(\gamma_1,\ldots,\gamma_p)$ and $\boldsymbol{\beta}=(\beta_1,\ldots,\beta_p)\in\R^p$, 
there is a symmetric quantum circuit $C$ that uses $O\bigl(pmn + m\log m\bigr)$ one-qubit and two-qubit gates and satisfies the following:
on every MAX 3XOR instance $J=(A,d)$ with $m$ equations and $n$ variables, encoded as $\ket{J}=\ket{A}\ket{d}$, its acceptance probability equals the probability that QAOA with angles $\boldsymbol{\gamma},\boldsymbol{\beta}$ produces an assignment satisfying at least $t$ equations.
\end{theorem}

This construction connects our lower bound methods to quantum optimization. The computation accepts when the QAOA algorithm produces an assignment satisfying at least $t$ equations (so its acceptance probability measures success in the optimization task). Our techniques can therefore be used to study the entanglement required by the full symmetric implementation, including its auxiliary qubits.

\subsection{Prior work}
Anderson and Dawar~\cite{Anderson2017} characterize fixed-point logic with counting by polynomially uniform families of symmetric circuits over a suitable gate set. Symmetry here is a property of the circuit structure: every permutation of the input labels respecting the encoding extends to an automorphism of the whole circuit. Inexpressibility in $\FPC$, as in the result of Cai, F\"urer, and Immerman~\cite{Cai1992}, consequently yields lower bounds for symmetric circuits~\cite{Dawar2015a}.

Dawar~\cite{Dawar2015} and Dawar and Wilsenach~\cite{Dawar2024a,Dawar2024} formulate the lower-bound argument directly using pebble games. Their method also applies to symmetric arithmetic circuits computing the permanent~\cite{Dawar2024a} and determinant~\cite{Dawar2024}. Dawar extends the CFI lower bound to other properties through symmetric circuit reductions~\cite{Dawar2015}. Our quantum indistinguishability theorem has the same role: it transfers a winning Duplicator strategy into equality of circuit output probabilities.

The symmetric quantum model of Castro-Silva, Gur, and Strelchuk supports amplitude amplification and phase estimation~\cite{CastroSilva2026}. They also give polynomial-size symmetric quantum circuits for XOR-SAT, obtaining an exponential separation from symmetric classical threshold circuits.
These upper bounds use arbitrary single-qubit gates and unbounded-fan-in threshold gates.
Using our symmetric implementation of threshold gates using 1-qubit and 2-qubit gates (Lemma~\ref{lem:symbqp-threshold}), we show that our models are equivalent in power.

Jozsa and Linden~\cite{Jozsa2003} show that polynomial-size uniform quantum computations over two-qubit gates can be simulated efficiently when, after every gate, their states factor into blocks of bounded size. Our result addresses the entanglement needed under a symmetry requirement, for a property that already has a classical polynomial-time algorithm. Note that we allow the block-size bound to grow to $n^{1-o(1)}$, thus going beyond the remit of their results.

\subsection{AI disclosure statement}
The definition of the quantum pebble game and the associated lower bounds were obtained and written without AI assistance. Theorems 1.4--1.6 were later obtained by Google Antigravity when prompted by the authors. The authors subsequently simplified and rewrote the proofs of those results.

\section{Circuit model, pebble games, and proof overview}
\label{sec:model}

Here, we formally describe our symmetric quantum circuit model. While our main results are stated for graphs, this model holds in greater generality for any finite relational structure, as given below.

\subsection{Preliminary definitions}

\begin{definition}
A \defn{relational vocabulary} is a nonempty finite tuple
$\tau=(R_1^{r_1},\ldots,R_k^{r_k})$ of relation symbols, where for each
$i\in[k]$, the relation symbol $R_i$ has an associated arity
$r_i\in\Z_+$.
\end{definition}

\begin{definition}
A \defn{$\tau$-structure} with respect to a relational vocabulary
$\tau=(R_1^{r_1},\ldots,R_k^{r_k})$ is a tuple
$\str{A}=\strb{A;R_1^{\str{A}},\ldots,R_k^{\str{A}}}$, where $A$ is a set called the
\defn{domain} of $\str{A}$ and for each $i\in[k]$, we have $R_i^{\str{A}}\subseteq A^{r_i}$.
\end{definition}

Let $\fin[\tau]$ denote the set of finite $\tau$-structures.
The elements of the domain $A$ are called the elements of $\str{A}$. The size
of a structure is the size of its domain. A structure $\str{A}$ is finite if its
domain is finite.

For example, consider the relational vocabulary $\taugraph=(E^2)$
consisting of one binary relation $E$. A directed graph can then be described
as a $\taugraph$-structure $\str{G}=\strb{V;E^{\str{G}}}$ where the
domain is the set of vertices $V$ and the edges are described by $E^{\str{G}}\subseteq V^2$.

A $\tau$-property is a subset of $\fin[\tau]$. A $\tau$-property ${\prop{P}}$ is said
to be polynomial time if there is a polynomial-time algorithm which, given an
efficient representation of a finite $\tau$-structure ${\str{A}}$, decides whether
${\str{A}}\in {\prop{P}}$. Say that a $\tau$-property ${\prop{P}}$ separates two structures
${\str{A}},{\str{B}}\in\fin[\tau]$ if ${\str{A}}$ and ${\str{B}}$ do not agree on membership in ${\prop{P}}$, and ${\prop{P}}$
separates two families of $\tau$-structures $({\str{A}}_n)_{n\in\Z_{\geq1}}$
and $({\str{B}}_n)_{n\in\Z_{\geq1}}$ if it separates ${\str{A}}_n$ and ${\str{B}}_n$ for each $n$.

For a finite set $S$, let $\Sym{S}$ denote the symmetric group of all permutations
on $S$. Let $\Alt{S}\leq\Sym{S}$ denote the alternating group of even permutations
on $S$. Write $\Sym{n}$ for $\Sym{[n]}$ and $\Alt{n}$ for $\Alt{[n]}$.
If $G$ is a group acting on $S$ and $s\in S$, let
$\Stab_G(s)=\{\sigma\in G:\sigma(s)=s\}$ denote the stabilizer group of $s$ and
$\Orb_G(s)=\{\sigma(s):\sigma\in G\}$ denote the orbit of $s$ under $G$.
If $T\subseteq S$, let
$\sigma(T)=\{\sigma(t):t\in T\}$ and let
$\Stab_G(T)=\bigcap_{t\in T}\Stab_G(t)$ denote the pointwise stabilizer of $T$.

Denote the group of unitaries on a Hilbert space $\Hilb$ by $\Unit(\Hilb)$.
If $\Hilb_S=(\C^2)^{\tprod S}$ and $G$ is a group acting on $S$,
there is a natural embedding $G\to\Unit(\Hilb_S)$ which maps
$\sigma\in G$ to the unitary on $\Hilb_S$ which permutes the qubits according
to the action of $\sigma$. Write $U_\sigma$ for this unitary, which is defined as
follows: if $\ket{\psi}=\ket{\psi_{s_1}}\tprod\cdots\tprod\ket{\psi_{s_N}}$
where $S=\{s_1,\ldots,s_N\}$, then
\begin{equation} \label{eqn:u-sigma-permutation}
U_\sigma\ket{\psi}
=\ket{\psi_{\sigma^{-1}(s_1)}}\tprod\cdots\tprod
\ket{\psi_{\sigma^{-1}(s_N)}}.
\end{equation}
That is, qubit $s\in S$ of a state $\ket{\psi}\in\Hilb_S$ corresponds to
qubit $\sigma(s)$ of $U_\sigma\ket{\psi}$: hence qubit $s$ of $U_\sigma\ket{\psi}$
is qubit $\sigma^{-1}(s)$ of $\ket{\psi}$.
Write $\Gamma_G=\{U_\sigma:\sigma\in G\}$ for the image of $G$ in
$\Unit(\Hilb_S)$ under this embedding.

Fix a relational vocabulary $\tau=(R_1^{r_1},\ldots,R_k^{r_k})$, and let
$\fin_n[\tau]$ denote the $\tau$-structures of size $n\in\Z_+$.
For simplicity, assume every structure in $\fin_n[\tau]$ has domain $[n]$.
We encode each structure ${\str{A}}\in\fin_n[\tau]$ as a bitstring $b({\str{A}})$ of length
$\ell_\tau(n)=\sum_{i=1}^k n^{r_i}$ as follows. Index the bitstring by the tuples
\begin{equation}
Q^\tau=\{(i,n_1,\ldots,n_{r_i}):i\in[k],\ n_1,\ldots,n_{r_i}\in[n]\},
\label{eq:qubit-labels}
\end{equation}
which we refer to as \defn{qubit labels}, and define the bit labelled by each
$(i,n_1,\ldots,n_{r_i})\in Q^\tau$ as
\begin{equation} \label{eqn:bitstring-repr}
b({\str{A}})_{(i,n_1,\ldots,n_{r_i})}=
\begin{cases}
1&\caseif (n_1,\ldots,n_{r_i})\in R_i^{\str{A}},\\
0&\caseif (n_1,\ldots,n_{r_i})\notin R_i^{\str{A}}.
\end{cases}
\end{equation}
This encoding forms a bijection between the bitstrings $\{0,1\}^{\ell_\tau(n)}$
and the structures in $\fin_n[\tau]$. Observe furthermore that $\ell_\tau(n)$ is
polynomial in $n$. We can then define
$\Hilb_n^\tau=(\C^2)^{\tprod Q^\tau}$,
and encode each structure ${\str{A}}\in\fin_n[\tau]$ as the computational basis state
$\ket{{\str{A}}}=\ket{b({\str{A}})}$. We often abbreviate $(i,n_1,\ldots,n_{r_i})$ to
$(i,\vec{n})$.

We refer to general states $\ket{A}\in\Hilb_n^\tau$ as \defn{quantum
$\tau$-structures}.

The action of a group $G$ on the domain $[n]$ of ${\str{A}}$ can be extended to an action
on $Q^\tau$ by letting $\sigma(i,\vec{n})=(i,\sigma(\vec{n}))$, where if
$\vec{n}=(n_1,\ldots,n_{r_i})$ then
$\sigma(\vec{n})=(\sigma(n_1),\ldots,\sigma(n_{r_i}))$.
$\Gamma_G\leq\Unit(\Hilb_n^\tau)$ is then the group of unitaries which
permutes the qubits according to $G$: we say $\Gamma_G$ is the \defn{classical
unitary group corresponding to $G$}.

\subsection{Symmetric quantum circuits}
\label{sec:symmetric-circuits}
We now define symmetric quantum circuits, following \cite{CastroSilva2026}.
A quantum circuit consists of a set of wires $W$ which define a Hilbert space
$\Hilb_W=(\C^2)^{\tprod W}$ upon which a sequence of unitary operators
called \defn{gates} act, chosen from a quantum gate set.
Our gate set $\gateset$ corresponds to all unitaries acting on at most two qubits.

\begin{remark}
Note that this formally differs from the gate set $\gateset_{thr}$ considered in \cite{CastroSilva2026}, which also includes threshold gates.
However, since 2-qubit gates can symmetrically implement threshold gates with only logarithmic overhead (see Lemma~\ref{lem:symbqp-threshold}), it easily follows from the arguments in that paper that our classes of symmetric circuits are equivalent up to a logarithmic factor in complexity.
\end{remark}
We define quantum circuits in terms of layers of commuting gates.
\begin{definition}[(Quantum circuit and layers)]
\label{def:quantum-circuit}
A quantum circuit over gate set $\gateset$ consists of a set of wires $W$ and a sequence of layers $(\layer_1,\ldots,\layer_m)$.
Each layer $\layer$ is a set of pairwise commuting unitary gates $U_g$ from $\gateset$ acting on $\Hilb_W$.
\end{definition}
To define a model of computation, we fix an output wire.
\begin{definition}[(Output $\tau$-circuit)]
\label{def:output-circuit}
An \defn{output quantum $\tau$-circuit of order $n$} is a quantum circuit whose
wires are partitioned as $W=X\sqcup A$, where
$X=\{x_{(i,\vec{n})}:(i,\vec{n})\in Q^\tau\}$ is the set of \defn{input wires}
and $A$ is the set of \defn{ancilla wires}, which has a distinguished
\defn{output wire} $w_{\out}\in W$.
\end{definition}

We often leave $\tau$ implicit for brevity. The Hilbert space of an output quantum
$\tau$-circuit of order $n$ is partitioned as
$\Hilb_W=\Hilb_n^\tau\tprod\Hilb_A$, where $\Hilb_n^\tau$ is
encoded on the input wires $X$ and the ancilla wires $A$ form the space
$\Hilb_A=(\C^2)^{\tprod A}$.
Such a circuit $C$ on layers $(\layer_1,\ldots,\layer_m)$ then defines a
function $\Hilb_n^\tau\to[0,1]$ as follows: we begin by initializing the
input wires $X$ to a quantum $\tau$-structure $\ket{\psi}\in\Hilb_n^\tau$
and each ancilla wire to $\ket{0}$, forming the state
$\ket{\psi}\ket{0}^{\tprod A}\in\Hilb_W$.
We then apply the unitaries from each layer in turn. For each layer
$\layer_i$, let $L_i$ denote the product of the unitaries in $\layer_i$
(in arbitrary order, since they commute). We say that $\layer_i$ is
\defn{represented} by the unitary $L_i$. We then form the state
\begin{equation}
\ket{\psi'}:=L_mL_{m-1}\cdots L_2L_1\ket{\psi}\ket{0}^{\tprod A}.
\end{equation}
Finally, we measure the wire $w_{\out}$ in the computational basis using
the observable
$O_{\out}:=\ket{1}_{w_{\out}}\bra{1}_{w_{\out}}
\tprod \iden_{W\setminus\{w_{\out}\}}$.
The probability of outcome $1$ is
 $C[\ket{\psi}]:=\braket[3]{\psi'}{O_{\out}}{\psi'}$.

Our model of quantum computation takes a quantum $\tau$-structure as input and
produces a classical probabilistic Boolean output. In our results, we will
typically be concerned only with inputs $\ket{{\str{A}}}$ which represent classical
$\tau$-structures ${\str{A}}$.
As before, a \defn{family of output quantum $\tau$-circuits $(C_n)_n$} consists of
one such circuit $C_n$ of order $n$ for each $n\in\Z_+$.

\begin{definition}[]
\label{def:computes-property}
A family of output quantum $\tau$-circuits $(C_n)_n$ \defn{computes} a
$\tau$-property ${\prop{P}}$ if for every $n\in\Z_+$ and
${\str{A}}\in\fin_n[\tau]$, we have $C_n[\ket{{\str{A}}}]\geq2/3$ if ${\str{A}}\in {\prop{P}}$ and
$C_n[\ket{{\str{A}}}]\leq1/3$ if ${\str{A}}\notin {\prop{P}}$.
\end{definition}

The \defn{size} $s(C)$ is the number of gates in an output quantum circuit $C$
and the \defn{number of ancillas} $a(C)$ is the size of its set $A$ of ancilla wires.

\begin{definition}[(Circuit automorphism)]
\label{def:circuit-automorphism}
An \defn{automorphism} of a quantum circuit $C$ with wires $W$ is a unitary
$V\in\Unit(\Hilb_W)$, which acts by permuting the wires $W$, such that for each layer $\layer$ of $C$
we have $\layer= V\layer V^\dagger:= \{VU_gV^\dagger:U_g\in\layer\}$.
\end{definition}

That is, an automorphism is a permutation of the wires which leaves each layer unchanged
up to an unimportant rearrangement of the gates. In other words, an automorphism
$V$ acts on the gates in each layer by conjugation. We also say that a unitary
$V\in\Unit(\Hilb_W)$ which acts by permuting the wires is an automorphism of a layer $\layer$ if
$\layer=\{VU_gV^\dagger:U_g\in\layer\}$, and of a unitary
$L\in\Unit(\Hilb_W)$ if $L=VLV^\dagger$ (that is, $V$ commutes with $L$).
Observe that if a unitary $L$ represents a layer $\layer$ and $V$ is an
automorphism of $\layer$, then $V$ is also an automorphism of $L$.
Given a permutation $\pi \in \Sym{W}$, we write $U_\pi \in U(\Hilb_W)$ for the unitary which permutes the wires according to $\pi$ as in equation (\ref{eqn:u-sigma-permutation}).

Consider an output quantum $\tau$-circuit $C$ of order $n$ with wires
$W=X\sqcup A$ and a symmetry group $\Gamma\leq\Unit(\Hilb_n^\tau)$.
A unitary $U\in\Gamma$ \defn{induces} an automorphism $V$
of $C$ if $V=U\tprod U_A$ for some unitary $U_A\in\Unit(\Hilb_A)$ which permutes the ancilla wires.
Note that the automorphism must act separately on the input wires and ancillas.

\begin{definition}[(Symmetric circuit)]
\label{def:permutation-symmetry}
Suppose $G\leq\Sym{n}$. Recall that $\Gamma_G\leq\Unit(\Hilb_n^\tau)$
is the unitary group which permutes the qubits according to the action of $G$.
An output quantum $\tau$-circuit of order $n$ is
\defn{$\Gamma_G$-symmetric} if every $U_\sigma\in\Gamma_G$ induces
an automorphism $V$ of $C$ which acts trivially on
$w_{\out}$, in the sense that
$V=\iden_{w_{\out}}\tprod V'_{W\setminus\{w_{\out}\}}$
for some $V'$ acting on the other qubits.
\end{definition}

That is, in a $\Gamma_G$-symmetric circuit, every permutation of the
input wires $U\in\Gamma_G$ induces a permutation of the ancillas $V'$ such that
$V=U\tprod V'$ is an automorphism fixing $w_{\out}$.
This is the permutation symmetry requirement of \cite{CastroSilva2026}, with the output wire fixed.

When working with $\Gamma_G$-symmetric circuits $C$ of order $n$, we also say that $\pi \in \Sym{W}$ is an automorphism of $C$ if $U_\pi$ %
is an automorphism of $C$, and we say $\sigma \in \Sym{n}$ induces an automorphism $\pi \in \Sym{W}$ if $U_\sigma \in \Unit(\Hilb^\tau_n)$ induces $U_\pi$.
Observe that if $\sigma$ induces $\pi$, then $\sigma$ and $\pi$ act identically on the input wires of $C$.

\subsection{Entanglement width}
We focus on quantum circuits taking classical structures as input, encoded as
computational basis states $\ket{{\str{A}}}$, and thereby obtain lower bounds on the
size of quantum circuits solving classical problems. Our technique is heavily
dependent on the amount of entanglement used by the circuit in the following sense.

\begin{definition}
\label{def:state-entanglement-width}
Let $\ket{\psi}\in(\C^2)^{\tprod n}$ be a state on $n$ qubits.
The \defn{entanglement width} of $\ket{\psi}$, denoted $\ew(\ket{\psi})$,
is the smallest integer such that $\ket{\psi}$ admits a decomposition
$\ket{\psi}=\ket{\psi_1}\tprod\cdots\tprod\ket{\psi_k}$, where each state
$\ket{\psi_i}$ is defined over a block $S_i$ of a partition $[n]=S_1\sqcup\cdots\sqcup S_k$, with
$\abs{S_i}\leq\ew(\ket{\psi})$.
\end{definition}

That is, if $\ew(\ket{\psi})\leq\ew$, then the state $\ket{\psi}$ uses at most
$\ew$-partite entanglement. Entanglement width was previously defined by Jozsa
and Linden \cite[Def.~4]{Jozsa2003}, where it was called being ``$p$-blocked''.

\begin{definition}
\label{def:circuit-entanglement-width}
Let $C$ be an output quantum $\tau$-circuit of order $n$ whose layers are
represented by unitaries $L_1,\ldots,L_m$.
Let $L_{\leq0}=\iden$ and $L_{\leq i}=L_iL_{i-1}\cdots L_1$ for each $i\in[m]$.
The \defn{entanglement width} of $C$ is defined as
\begin{equation}
\ew(C):=\max_{\substack{0\leq i\leq m\\\str{A}\in\fin_n[\tau]}}
\ew\bigl(L_{\leq i}\ket{{\str{A}}}\ket{0}^{\tprod a(C)}\bigr).
\label{eq:entanglement-width}
\end{equation}
\end{definition}

That is, if $\ew(C)\leq\ew$, then when a classical structure $\ket{{\str{A}}}$
(\ie, a computational basis state) is given as input, after every layer, the state
of the circuit uses at most $\ew$-partite entanglement.
States $\ket{\psi}$ with $\ew(\ket{\psi})=1$ are product states, and circuits
$C$ with $\ew(C)=1$ operate (between layers) only on product states.
The block partition may change between layers and between inputs. Including layer $0$ gives width $1$ to a circuit with no layers, since the initial state is a product state. The maximum in Definition~\ref{def:circuit-entanglement-width} is taken over basis inputs. It does not impose a bound on computations starting from arbitrary quantum superpositions of structures.

\subsection{Partial trace-based pebble games}
\label{subsec:pebble-games}
A pebble records a constraint on the allowed relabellings of the input domain. In the classical game, Spoiler tries to expose a difference between two structures using a bounded number of such constraints, while Duplicator chooses compatible relabellings. In the quantum game, the comparison is made through the reduced density matrices on the qubits fixed by the current constraints. The number of pebbles counts domain elements associated with the considered constraints, whereas the partial trace acts on input qubits.

Denote the set of density matrices on $\Hilb$ by $\Dens(\Hilb)$.
If $\Hilb=(\C^2)^{\tprod X}$ is composed of $\abs{X}$ qubits labelled by
elements of a finite set $X$ and $S\subseteq X$ is nonempty, let $\Tr_S$ denote
the partial trace operation which ``traces out'' the qubits $S$.
Additionally, let
\begin{equation}
\TrTo{S}=\Tr_{X\setminus S}:
\Dens((\C^2)^{\tprod X})\to
\Dens((\C^2)^{\tprod S})
\end{equation}
denote the partial trace operation which ``traces out'' the other qubits,
leaving only the qubits $S$.
If $x\in X$ is a single qubit, write $\TrTo{x}$ for
$\TrTo{\{x\}}$.

\begin{definition}[(Pebble set)]
\label{def:pebble-set}
Given a group $\Gamma\leq\Unit(\Hilb_n^\tau)$, a \defn{pebble set}
$\Peb$ for $\Gamma$ is a set of subgroups of $\Gamma$, called
\defn{pebble groups} or simply \defn{pebbles} and denoted $\Gamma_p\in\Peb$,
obeying the following properties:
\begin{enumerate}[label=(\alph*)]
\item $\Peb$ is closed under conjugation by $\Gamma$: for every
$\Gamma_p\in\Peb$ and every $U\in\Gamma$, $U\Gamma_pU^\dagger\in\Peb$.
\item Each pebble group $\Gamma_p\in\Peb$ is a self-normalizing subgroup
of $\Gamma$: for every $U\in\Gamma$, if $U\Gamma_pU^\dagger=\Gamma_p$ then
$U\in\Gamma_p$.
\end{enumerate}
\end{definition}

Write $\inter P:=\bigcap_{\Gamma_p\in P}\Gamma_p$ for the intersection of the
pebble groups in $P$, which is just the group of symmetries respecting all the
pebbles in $P$. By convention, we set $\inter\varnothing:=\Gamma$.
We say that a permutation group $G\leq\Sym{n}$ has \defn{distinct stabilizers} if for every
$a,b\in[n]$ with $a\neq b$, we have $\Stab_G(a)\neq\Stab_G(b)$.

\begin{definition}[(Classical pebble set)]
\label{def:classical-pebble-set}
Suppose $\Gamma_G\leq\Unit(\Hilb_n^\tau)$ is the classical unitary group corresponding to a group $G\leq\Sym{n}$ which has distinct stabilizers,
and let $\phi:G\to\Gamma_G$ be the isomorphism mapping $\sigma\mapsto U_\sigma$.
The \defn{classical pebble set} for $\Gamma_G$ is
\begin{equation}
\Peb_G=\{\phi(\Stab_G(a)):a\in[n]\}.
\end{equation}
\end{definition}

That is, the classical pebble groups in $\Peb_G$ are just the subgroups
which fix each domain element.
The condition that $G$ has distinct stabilizers is sufficient to show that each pebble in $\Peb_G$ is a self-normalizing subgroup of $\Gamma_G$: if $\phi(\Stab_G(a)) = U_\sigma\phi(\Stab_G(a))U_\sigma^\dagger$, then $\Stab_G(a) = \sigma\Stab_G(a)\sigma^{-1} = \Stab_G(\sigma a)$, so the distinct stabilizers property tells us that $\sigma a = a$, i.e. $\sigma \in \Stab_G(a)$.

For each qubit label $(i,\vec{n})\in Q^\tau$, write
$\Stab_\Gamma(i,\vec{n})$ for the subgroup of $\Gamma$ which acts trivially
on that qubit. In particular, if $\Gamma_G$ is a classical unitary group and $\phi$ is the natural embedding
of $G$ into $\Gamma_G$, then
\[
\Stab_{\Gamma_G}(i,\vec{n})=\phi(\Stab_G((i,\vec{n}))).
\]

\begin{definition}[\cite{Dawar2024}]
\label{def:pebble-lift}
Let $P\subseteq\Peb$ be a set of pebbles. The \defn{lift} of $P$ is the
set of qubits
\begin{equation}
Q_P^\tau=\{(i,\vec{n})\in Q^\tau:\inter P\leq\Stab_\Gamma(i,\vec{n})\}.
\end{equation}
\end{definition}

That is, $Q_P^\tau$ is the set of qubit labels which are pointwise stabilized
by the pebbles. This definition of lift is similar to that given by Dawar and
Wilsenach \cite{Dawar2024}.

\begin{definition}
\label{def:ell-pt-isomorphism}
A unitary $U\in\Gamma$ is an \defn{$\ell$-partial trace-based partial isomorphism},
abbreviated \defn{$\ell$-PT partial isomorphism}, from $\ket{A}$ to $\ket{B}$
with respect to $P\subseteq\Peb$ if for every $S\subseteq Q_P^\tau$ with
$1\leq\abs{S}\leq\ell$ we have
\begin{equation}
\TrTo{S}(\ketbra{A}{A})
=\TrTo{S}(U^\dagger\ketbra{B}{B}U).
\end{equation}
\end{definition}

\begin{definition}
\label{def:pt-isomorphism}
A unitary $U\in\Gamma$ is a \defn{partial trace-based partial isomorphism},
abbreviated \defn{PT partial isomorphism}, from $\ket{A}$ to $\ket{B}$ with
respect to $P\subseteq\Peb$ if
\begin{equation}
\TrTo{Q_P^\tau}(\ketbra{A}{A})
=\TrTo{Q_P^\tau}(U^\dagger\ketbra{B}{B}U).
\end{equation}
\end{definition}

\begin{definition}
\label{def:pt-game}
The \defn{$(\Gamma,\Peb,k)$-partial trace-based quantum pebble game},
abbreviated as the \defn{$(\Gamma,\Peb,k)$-PT quantum pebble game}, on
$\ket{A}$ and $\ket{B}$ is defined as follows.
The state of the game consists of a subset of pebbles $P\subseteq\Peb$
with $\abs{P}\leq k$ and a unitary $U\in\Gamma$. Initially, no pebbles are placed
and $U=\iden$. Each round proceeds as follows.
\begin{enumerate}
\item Optionally, Spoiler removes a pebble from $P$. If $\abs{P}=k$, Spoiler must
remove a pebble.
\item Duplicator chooses a unitary $V\in\inter P$ and sets $U:=UV$.
\item Spoiler chooses a pebble from $\Peb\setminus P$ and adds it to $P$.
\end{enumerate}
Spoiler wins if $U$ is not a PT partial isomorphism from $\ket{A}$ to
$\ket{B}$ with respect to $P$. Otherwise, play continues, and Duplicator
wins if it has a strategy to play forever.
The \defn{$(\Gamma,\Peb,k)$-$\ell$-PT quantum pebble game} is defined
identically with the notion of PT partial isomorphism replaced by
$\ell$-PT partial isomorphism.
\end{definition}

Clearly if Duplicator wins the $(\Gamma,\Peb,k)$-PT quantum pebble game,
then it also wins the $(\Gamma,\Peb,k)$-$\ell$-PT quantum pebble game for
every $\ell$. The PT quantum pebble game provides a simple and natural quantum
generalization of Dawar and Wilsenach's $(G,k)$-pebble game from \cite{Dawar2024}.

\subsection{The lower-bound technique}
\label{sec:proof-overview}

The lower bound combines small supports with inputs that require many pebbles to distinguish. A support is a set of input vertices such that fixing those vertices fixes the gate or wire under consideration. The classical argument applies this idea to gates. For a quantum circuit, the reduced state of one wire need not determine the effect of a later gate, because that wire can be entangled with other wires. We instead use reduced density matrices on tuples of wires, with the entanglement width determining the tuple sizes needed for reconstruction.

There are three steps. First, bounds on circuit size give bounds on the orbits of wire tuples and hence on their support sizes. Second, reconstruction and the action of each commuting layer yield an indistinguishability theorem: if the required supports have size at most $k$, a circuit cannot distinguish inputs on which Duplicator wins with $2k$ pebbles. Third, the CFI construction supplies pairs that remain indistinguishable with linearly many pebbles. Since $\PAltCFI$ separates those pairs, any circuit computing it must violate the combined size and entanglement restrictions.

The conclusion of the indistinguishability theorem is equality of output probabilities. Thus the argument excludes any nonzero distinguishing bias on a hard pair whenever its hypotheses hold. The error bound $1/3$ is the convention used to define computation of a property.

We apply essentially the same technique abstractly to tuples of distinct
wires. Fix a set of wires $W$ and $\ell\in\Z_+$ and assume
$\abs{W}\geq\ell$, and define
\[
\Wd{\ell}=\{(w_1,\ldots,w_\ell)\in W^\ell:
w_1,\ldots,w_\ell\text{ are all distinct}\}.
\]
Any group $G$ acting on $W$, in particular $\Sym{W}$, then also acts on
$\Wd{\ell}$ elementwise.

Fix a symmetry group $G\leq\Sym{n}$ and let $C$ be a
$\Gamma_G$-symmetric output quantum circuit of order $n$.
Suppose $C$ has wires $W=X\sqcup A$, where $X$ are the input wires and
$A$ are the ancilla wires, and let $w_{\out}$ be its output wire.
Let $\Aut(C)\leq\Sym{W}$ be the group of automorphisms of $C$, \ie,
the group of permutations of the wires which map the input wires $X$ to
$X$, fix $w_{\out}$, and induce an automorphism of the circuit.
Let $\In(C)=\Stab_{\Aut(C)}(X)$ be the group of internal automorphisms
of $C$, which are the automorphisms that pointwise fix each input wire
$x\in X$. Observe that $\In(C)$ is a normal subgroup of $\Aut(C)$.

An input permutation need not have a unique extension to the circuit wires. Any two extensions differ by an internal automorphism. Passing from wire tuples to their orbits under $\In(C)$ makes the induced action independent of this choice. For each $\ell$-size tuple of distinct wires $\vec{w}\in \Wd{\ell}$, define its orbit under $\In(C)$ as
\[
[\vec{w}]:=\{\pi\vec{w}:\pi\in\In(C)\},
\]
where we recall that $\Aut(C)$ acts elementwise on tuples of wires.
The collection of these orbits is denoted $\Wd{\ell}/\In(C)$.
Observe that $\Aut(C)/\In(C)$, and hence $G$, acts on
$\Wd{\ell}/\In(C)$ as $[\pi][\vec{w}]:=[\pi\vec{w}]$, where
$[\pi]=\pi\In(C)$.

Explicitly, the action of $\sigma\in G$ takes $[\vec{w}]$ to
$[\pi_\sigma\vec{w}]$, where $\pi_\sigma$ is an automorphism
induced by $\sigma$.
A support for $[\vec{w}]$ is a set of input vertices such that fixing those vertices fixes the entire orbit $[\vec{w}]$. It need not fix each representative tuple individually. Small supports allow the relevant tuples to be described using few pebbles. Definition~\ref{def:tuple-support} gives the group-theoretic formulation.

We define a ``directional partial trace'' $\DTrTo{\vec{w}}(\rho)$ of a density matrix
$\rho\in\Dens(\Hilb_W)$ with respect to a tuple
$\vec{w}\in \Wd{\ell}$ which essentially traces out the qubits not in
$\vec{w}$ and arranges the remaining qubits in the order specified by
$\vec{w}$. This ordering allows reduced states on different tuples of the same length to be compared as density matrices on the same tensor-product space. The reconstruction argument in Section~\ref{sec:reconstruction} relates matching collections of these reduced states to permutations of the underlying wires.

For the indistinguishability theorem, we prove the contrapositive: assume
$C[\ket{\str{A}}]\neq C[\ket{\str{B}}]$.
We give a Spoiler strategy. Let the layers of $C$ be represented by
unitaries $L_1,\ldots,L_m$, let $L_{\leq i}=L_iL_{i-1}\cdots L_1$, with $L_{\leq0}=\iden$, and let
\[
\begin{aligned}
\rho_i^{\str{A}}
&=L_{\leq i}\bigl(\ketbra{\str{A}}{\str{A}}
\tprod\ketbra{0}{0}^{\tprod A}\bigr)L_{\leq i}^{\dagger}
\\
\rho_i^{\str{B}}
&=L_{\leq i}\bigl(\ketbra{\str{B}}{\str{B}}
\tprod\ketbra{0}{0}^{\tprod A}\bigr)L_{\leq i}^{\dagger}.
\end{aligned}
\]
For each $i$ from $m$ to $0$, Spoiler plays to maintain the invariant
that there is some $\vec{w}_i\in \Wd{c_i}$ for some $c_i\leq\ell$
such that $[\vec{w}_i]$ is supported by the current set of pebbles $P$
and for every automorphism $\pi$ induced by the current $\sigma\in G$,
we have
\[
\DTrTo{\vec{w}_i}(\rho_i^{\str{A}})
\neq\DTrTo{\vec{w}_i}(U_\pi^\dagger\rho_i^{\str{B}}U_\pi).
\]
Different output probabilities provide the initial discrepancy on the output wire. The partition-symmetry argument transfers a discrepancy backwards through each original layer, using tuples of at most $\ell=\ew(C)$ wires. Supports translate these tuples into pebble positions. At the initial layer, the remaining discrepancy contradicts the winning strategy of Duplicator on the encoded classical inputs. Sections~\ref{sec:reconstruction} and~\ref{sec:indistinguishability} contain the corresponding technical arguments.

\subsection{Symmetric simulation without an entanglement restriction}

We now give an overview of the proof technique for Theorem~\ref{thm:symbqp-simulation}, which shows that polynomial-time uniform families of symmetric quantum circuits without a restriction on entanglement width can compute every graph property in BQP. To prove this, we simulate an arbitrary quantum algorithm which computes a graph property. The main idea is to satisfy the symmetry requirement by running the original algorithm on all orderings of the vertices in superposition. To do so, we create an auxiliary register which represents a map $\lambda : [n] \to \{0,\dots,M-1\}$, where $n$ is the number of vertices in the input graph and $M$ is a power of two chosen to be large enough that such a map $\lambda$ chosen uniformly at random is likely to be injective. We initialize the register in a uniform superposition of all such maps: $\frac{1}{\sqrt{M^n}} \sum_{\lambda : [n] \to \{0,\dots,M-1\}} \ket\lambda$.
A symmetric Boolean circuit sorts the vertices according to their values under $\lambda$ and writes the input graph in the resulting order, provided these values are all distinct. Among the branches of the superposition where this condition holds, every ordering of the vertices occurs equally often.

Every permutation of the input vertices extends to a permutation of the circuit wires that fixes the output wires of the sorting gadget. We can apply the original quantum algorithm on these wires, with any additional workspace wires also fixed. The sorting gadget produces the input graph with its vertices reordered whenever $\lambda$ is injective. Otherwise, the computation may give an incorrect answer. Since we retain the label register, the acceptance probability is the average over the choices of $\lambda$, so the additional error is at most the probability that $\lambda$ is not injective. We first reduce the error of the original algorithm by a constant number of repetitions, leaving room for this additional error. Choosing $M$ sufficiently large relative to $n$ then ensures that the overall computation satisfies the required bounded-error guarantee.

The sorting gadget is expressed in terms of threshold gates, which compute whether the Hamming weight of the input is at least a fixed threshold $t$. We show that threshold gates have an efficient symmetric implementation using one-qubit and two-qubit gates. The construction first encodes the Hamming weight in the phases of a counter register, then applies an inverse quantum Fourier transform to recover its binary representation. We compare this value with $t$ and uncompute the workspace while retaining the result of the comparison. The technical arguments are in Section~\ref{sec:symbqp}.

This result shows that the classical exponential lower bound on the size of symmetric circuits computing the hard property does not directly extend to the quantum setting: indeed, Theorem~\ref{thm:symbqp-simulation} implies that $\PAltCFI$ can be computed by polynomial-size symmetric quantum circuits. A superpolynomial size lower bound for symmetric quantum circuits deciding this property therefore requires an additional restriction. Theorem~\ref{thm:main} gives such a lower bound when entanglement width is restricted.

\subsection{Symmetric simulation without entanglement}

Theorem~\ref{thm:symbqp-simulation} shows that without a restriction on entanglement width, symmetric quantum circuits can compute every graph property in BQP. Theorem~\ref{thm:ew-1-computes-fpc} shows that even without entanglement, they can compute every graph property definable in $\FPC$. Here, the absence of entanglement means that the state is a product of single-qubit states at every layer boundary on computational-basis inputs (the condition of entanglement width being $1$). Our proof uses Anderson and Dawar's characterization of $\FPC$ in terms of classical symmetric circuits with threshold gates~\cite{Anderson2017}. We simulate these circuits using one-qubit and two-qubit gates while preserving this product structure.

\begin{theorem}[{\cite{Anderson2017}}]
\label{thm:anderson-dawar-fpc}
    A graph property is computed by a polynomial-time uniform family of symmetric classical circuits using threshold and NOT gates if and only if it is definable in $\FPC$.
\end{theorem}

Hence, we show that symmetric quantum circuits without entanglement can efficiently simulate any symmetric classical circuit using threshold and NOT gates.
Our threshold gate construction described above does not use entanglement if the input is in the computational basis, so this simply requires translating from the DAG-based classical circuit model to reversible circuits.
A simple algorithm for doing so was given in \cite[Prop.~1]{CastroSilva2026}.
The technical discussion appears at the end of Section~\ref{sec:symbqp}.

\subsection{Symmetric implementation of QAOA}
\label{subsec:qaoa-overview}

Finally, we outline the construction for Theorem~\ref{thm:qaoa-compilation}. The input qubits store the equations of the MAX 3XOR instance. We use a separate set of qubits, one for each variable, to prepare the QAOA state. To implement each cost operation, we compute whether each equation is satisfied, apply the corresponding phase, and undo the computation. The mixer applies the same single-qubit operation to every assignment wire.

After the final round, we again compute which equations are satisfied. The symmetric threshold computation of Lemma~\ref{lem:symbqp-threshold} then records whether their number is at least $t$, without measuring the assignment. This uses the same counting construction as the preceding simulation results.

The product and parity wires follow the equation and variable labels under permutation. The counter, comparison, and output wires are fixed. The gates can be arranged into commuting layers preserved by these permutations, using only one-qubit and two-qubit gates. The output probability is therefore the probability that QAOA produces an assignment meeting the threshold. Appendix~\ref{app:qaoa} gives the construction and gate count.

\section{Discussion and open questions}

One might wonder whether imposing sub-linear entanglement width is overly restrictive in the presence of permutation symmetry. Indeed, a well-known property of permutation-invariant pure states is that they are either fully separable or genuinely multipartite entangled~\cite{ichikawa2008exchange}. However, this dichotomy does not apply to our computational model:
vertex permutations may induce nontrivial permutations of the workspace qubits, allowing the circuit to maintain multiple entangled blocks that are themselves permuted by the symmetry group.
It is not hard to construct examples that illustrate this.

In particular, symmetric quantum circuits of entanglement width at most $n^{1-\epsilon}$ need not be restricted to separable states or trivial quantum dynamics. They can generate and manipulate genuinely multipartite entanglement involving a polynomially growing number of qubits, while remaining equivariant under all vertex permutations. More generally, the restriction allows nontrivial quantum computations within entangled blocks, whose size may grow polynomially with $n$. Establishing lower bounds in this setting therefore requires exploiting the interplay between the symmetry of the computation and the limitations imposed on its entanglement structure, rather than either restriction in isolation.

Our results raise fundamental questions about the power of symmetric quantum computation, the role of entanglement in determining that power, and the extent to which it can be characterized by quantum pebble games.

\begin{enumerate}

\item \textbf{Is linear entanglement width necessary?} Corollary~\ref{cor:poly} rules out polynomial-size symmetric circuits of entanglement width $o(n/\log n)$ for $\PAltCFI$. Can this bound be strengthened to linear width? More precisely, for each fixed $d>0$, does there exist $c_d>0$ such that, for all sufficiently large Alt-CFI pairs on $n$ vertices from Section~\ref{sec:cfi}, every symmetric circuit $C$ with $s(C)\leq n^d$ and different acceptance probabilities on the two graphs reaches a state of entanglement width at least $c_dn$ on each input?

Here we allow arbitrary ancillary qubits and do not require a unique extension of an input permutation to the circuit. In particular, internal automorphisms can permute many ancillary wires while fixing every input wire. Can this freedom reduce the entanglement needed to distinguish the hard pairs, or is linear width necessary even in its presence?

\item \textbf{How does the quantum pebble game compare with Weisfeiler--Leman refinement?} On computational-basis encodings of graphs, our partial-trace game recovers the classical pebble game underlying the Weisfeiler--Leman hierarchy. How does this correspondence change when graphs are encoded as entangled quantum states? In particular, does comparing their reduced density matrices reveal distinctions that classical refinement misses, or conceal distinctions it detects, at comparable numbers of pebbles?

Consider the $\ell$-PT quantum pebble game of Section~\ref{subsec:pebble-games} on general quantum inputs, with $k$ pebbles. At each position, the game compares reduced density matrices on at most $\ell$ qubits. Can this notion of indistinguishability be characterized by a refinement procedure analogous to classical Weisfeiler--Leman refinement? How does it relate to the distinguishing power of symmetric quantum circuits with restricted entanglement?

\item We conjecture that \textbf{symmetric quantum circuits of any fixed entanglement width have the power of $\FPC$}. More precisely, fix an integer $b\geq1$. Every graph property computed with bounded error by a polynomial-time uniform family $(C_n)_n$ of polynomial-size symmetric quantum circuits with $\eta(C_n)\leq b$ is definable in $\FPC$. The circuits use one-qubit and two-qubit gates, whose entries are uniformly approximable to $r$ bits of precision in time polynomial in $n$ and $r$. No further restriction is imposed on depth, and the partition into entangled blocks may depend on both the input and the layer.

Theorem~\ref{thm:ew-1-computes-fpc} proves the converse already at width $1$. The conjecture asserts that allowing entangled blocks of any fixed size does not enlarge the class of graph properties that can be computed.
Together with Theorem~\ref{thm:symbqp-simulation}, this would identify a boundary between the computational power available with bounded and unbounded entanglement width: every graph property in BQP admits an efficient symmetric implementation, but deciding a property beyond $\FPC$ requires unbounded entanglement width.

\end{enumerate}

\section{Internal automorphisms and partition symmetry}\label{sec:reconstruction}

\subsection{Properties of internal automorphisms}

We begin by characterizing the automorphisms of the circuit $C$ in terms of $G$ and $\In(C)$.

\begin{proposition} \label{prop:embedding}
Let $G \le \Sym{n}$ and let $C$ be a $\Gamma_G$-symmetric output quantum circuit.
There is an embedding $\chi : G \to \Aut(C)/\In(C)$ given by $\sigma \mapsto [\pi_\sigma] = \pi_\sigma \In(C)$ which maps $\sigma \in G$ to exactly the set of automorphisms induced by $\sigma$,
where $\pi_\sigma$ is any automorphism induced by $\sigma$.
\end{proposition}
\begin{proof}
Any $\pi \in \Aut(C)$ induced by $\sigma$ acts identically to $\sigma$, and so $\pi_\sigma$, on the input wires $X$ of $C$, so $\pi_\sigma^{-1}\pi \in \In(C)$ and so $\pi = \pi_\sigma(\pi_\sigma^{-1}\pi) \in \pi_\sigma\In(C)$.
Conversely, for every $\pi' \in \In(C)$, $\sigma$ induces $\pi_\sigma\pi'$ as it acts identically to $\pi_\sigma$ on $X$.
So $\sigma$ induces exactly $[\pi_\sigma]$.
For an embedding, suppose $\chi(\sigma_1) = \chi(\sigma_2)$: then $\sigma_1$ and $\sigma_2$ both induce some $\pi \in \Aut(C)$, so they must act the same on $X$.
The action of $G$ on $X$ is faithful since $X$ is nonempty, so $\sigma_1 = \sigma_2$.
\end{proof}

This justifies that $\sigma \in G$ acts on $\Wd\ell/\In(C)$ by taking $[\vec w]$ to $[\pi_\sigma \vec w]$.
If $H \le G$, we write $\chi(H) = \bigcup_{\sigma \in H} \chi(\sigma) \le \Aut(C)$ for the group of automorphisms induced by elements of $H$.

The crucial property of internal automorphisms is that the state of $C$ between each layer is symmetric against each internal automorphism, regardless of the input state.
Formally, we have the following. %

\begin{proposition} \label{prop:internal-aut-stabilizes}
Let $G \le \Sym{n}$ and let $C$ be a $\Gamma_G$-symmetric output quantum circuit with layers represented by unitaries $L_1,\dots,L_m$.
Let $L_{\le 0} = \iden$ and $L_{\le i} = L_iL_{i-1}\cdots L_1$ for each $i \in [m]$.
Let $\ket\psi \in \Hilb^\tau_n$ be any input state: then $\ket{\psi_i} = L_{\le i}\ket{\psi}\ket{0}^{\tprod a(C)}$ is the state of $C$ after any layer $0 \le i \le m$.
For every internal automorphism $\pi \in \In(C)$, we have $U_\pi \ket{\psi_i} = \ket{\psi_i}$.
\end{proposition}
\begin{proof}
Since $\pi$ is an automorphism of $C$, $U_\pi$ commutes with every layer, and since $\pi$ does not move any of the input wires, $U_\pi\ket\psi\ket0^{\tprod a(C)} = \ket\psi\ket0^{\tprod a(C)}$.
\end{proof}

We work with equivalence classes of wires $[\vec w] \in \Wd\ell/\In(C)$.
This formalism has the following useful property:

\begin{lemma} \label{lem:g-to-aut-stabilizes} %
Let $\vec w \in \Wd\ell$.
Every $\sigma \in G$ which stabilizes $[\vec w]$ induces some $\pi \in \Aut(C)$ which stabilizes $\vec w$.
\end{lemma}
\begin{proof}
$\In(C)$ acts transitively on $[\vec w]$ by definition.
So, suppose $\sigma$ induces some $\pi_\sigma \in \Aut(C)$ which maps $\vec w$ to $\vec v \in [\vec w]$.
There is some $\pi' \in \In(C)$ mapping $\vec v$ to $\vec w$, and by Proposition~\ref{prop:embedding}, $\sigma$ also induces $\pi'\pi_\sigma$.
\end{proof}

\subsection{Support and support theorem} %

Our lower bound technique, like that of Dawar and Wilsenach, relies on the notion of support and a support theorem which translates bounds on the size of orbits of circuit elements into bounds on their support size.
In our setting, the circuit elements under consideration are equivalence classes of tuples of wires $[\vec w] \in \Wd\ell/\In(C)$, so we generalize the corresponding definitions as follows.
We identify $G$ with $\Gamma_G$ through $\sigma\mapsto U_\sigma$.

\begin{definition}
\label{def:tuple-support}
A set of pebbles $P\subseteq\Peb$ \defn{supports} an orbit
$[\vec{w}]\in \Wd{\ell}/\In(C)$ if
$\inter P\leq\Stab_G([\vec{w}])$.
\end{definition}

\begin{definition}
\label{def:maximum-tuple-support}
Let $\minsupp([\vec{w}])$ be the size of the smallest set
of pebbles $P\subseteq\Peb$ which supports $[\vec{w}]$. Let
\begin{equation}
\maxsupp_\ell(C)
:=\max_{[\vec{w}]\in \Wd{\ell}/\In(C)}
\minsupp([\vec{w}]).
\end{equation}
\end{definition}

\begin{definition}
\label{def:maximum-tuple-orbit}
Let
\begin{equation}
\maxorb_\ell(C)
:=\max_{[\vec{w}]\in \Wd{\ell}/\In(C)}
\abs{\Orb_{\Aut(C)/\In(C)}([\vec{w}])}.
\end{equation}
\end{definition}

As in Dawar and Wilsenach's technique \cite{Dawar2024}, our support theorem is based on the following group-theoretic lemma.

\begin{lemma}[{\cite[Thm. 5.2B]{Dixon1996}}] \label{lem:altn-support-lemma}
    Suppose that $n > 8$ is an integer, and let $k$ be an integer such that $1 \le k \le n/4$.
    Suppose that $H \le \Alt{n}$ has index $[\Alt{n} : H] < \binom{n}{k}$.
    There exists some $S \subseteq [n]$ with $\abs{S} < k$ such that $\Stab_{\Alt{n}}(S) \le H$.
\end{lemma}

Our support theorem is below.
It gives an explicitly quantitative rather than asymptotic bound on the parameters defined above, but otherwise closely follows the proof of Dawar and Wilsenach's support theorem \cite[Thm.\ 4.11]{Dawar2024}.

\begin{theorem}[(Quantum support theorem)] \label{thm:support}
    Let $(C_n)_n$ be a family of $\Gamma_{\Alt{n}}$-symmetric output quantum circuits and let $\ell \in \Z_+$.
    For all $n > 8$, if $\maxorb_\ell(C_n) \le 2^{n/4-1}$, then $\maxsupp_\ell(C_n) \le \log_2 \maxorb_\ell(C_n)$.
\end{theorem}
\begin{proof}
    For brevity, write $b(n) = \maxorb_\ell(C_n)$.
    Let $k$ be the smallest positive integer such that $b(n) < \binom{n}{k}$: we have $\binom{n}{k-1} \le b(n) < \binom{n}{k}$.
    Observe that $k \le n/2$, since otherwise $\binom{n}{k-1} \ge \binom{n}{k}$.
    Since $\binom{n}{m} \ge (\frac{n}{m})^m$ whenever $1 \le m \le n$, if $k \ge 2$, then
    \[ b(n) \ge \binom{n}{k-1} \ge \left(\frac{n}{k-1}\right)^{k-1} \ge 2^{k-1}. \]
    If $k = 1$, we also have $b(n) \ge 2^{k-1} = 1$ automatically.
    Hence $k \le \log_2 b(n) + 1$.
    By the hypothesis, $k \le n/4$.

    For every $[\vec w] \in \Wd\ell/\In(C)$, we have $\abs{\Orb_{\Aut(C_n)/\In(C_n)}([\vec w])} < \binom{n}{k}$, and so since $\Alt{n}$ embeds into $\Aut(C_n)/\In(C_n)$, $\abs{\Orb_{\Alt{n}}([\vec w])} < \binom{n}{k}$.
    By the orbit--stabilizer theorem, then, $[\Alt{n} : \Stab_{\Alt{n}}([\vec w])] < \binom{n}{k}$, so by Lemma~\ref{lem:altn-support-lemma}, there is some $S \subseteq [n]$ with $\abs{S} < k$ such that $\Stab_{\Alt{n}}(S) \le \Stab_{\Alt{n}}([\vec w])$.
    Hence the set of pebbles $P = \{\Stab_{\Alt{n}}(s) : s \in S\} \subseteq \Peb_{\Alt{n}}$ supports $[\vec w]$ and has size less than $k \le \log_2 b(n) + 1$.
    So $\maxsupp_\ell(C_n) \le \log_2 b(n)$.
\end{proof}

\subsection{The \texorpdfstring{$\ell$}{l}-part function and partition symmetry} %

The directional partial trace is formally defined as follows.
Fix some total order $\le$ on $W$ and view $\Hilb_W = (\C^2)^{\tprod W}$.
For each $S \subseteq W$ with $\abs{S} = \ell$, we view $\TrTo{S}$ as a function $\Dens(\Hilb_W) \to \Dens((\C^2)^{\tprod \ell})$, where for each $\rho \in \Dens(\Hilb_W)$, the qubits of $\TrTo{S}(\rho)$ are indexed $1,\dots,\ell$ and correspond to the qubits of $\rho$ indexed by the wires $S$ arranged in ascending order according to $\le$.

\begin{definition}
Let $\vec w = (w_1,\dots,w_\ell) \in \Wd\ell$ and $\rho \in \Dens(\Hilb_W)$.
Let $\pi \in \Sym{\ell}$ be the unique permutation such that $(w_{\pi(1)},\dots,w_{\pi(\ell)})$ is in ascending order according to $\le$.
Let $S = \{w_1,\dots,w_\ell\}$.
The \defn{directional partial trace} of $\rho$ with respect to $\vec w$ is
\begin{equation}
\DTrTo{\vec w}(\rho) := U_\pi \TrTo{S}(\rho) U_\pi^\dagger.
\end{equation}
\end{definition}

The $i$th qubit of $\DTrTo{\vec w}(\rho)$ corresponds to the $w_i$ qubit of $\rho$, and in particular does not depend on the total order chosen.
Observe that if $\pi \in \Sym{W}$ acts on $\Wd\ell$ elementwise, then
\begin{equation} \label{eqn:dtr-property}
\DTrTo{\pi\vec w}(\rho) = \DTrTo{\vec w}(U_\pi^\dagger \rho U_\pi).
\end{equation}

In our lower bound technique, we look at $\ell$ wires of the circuit at a time.
This notion is captured via the $\ell$-part function, which is the central piece of data used in our argument.

\begin{definition}
For every density matrix $\rho \in \Dens(\Hilb_W)$ and $\ell \in \Z_+$, its \defn{$\ell$-part function} $f_\rho : \Wd\ell \to \Dens((\C^2)^{\tprod \ell})$
is defined as \[ f_\rho(\vec w) = \DTr_{\vec w}(\rho). \]
For $S \subseteq \Wd\ell$, write $f^S_\rho : S \to \Dens((\C^2)^{\tprod\ell})$ for the restriction of $f_\rho$ to the domain $S$.
\end{definition}

By equation (\ref{eqn:dtr-property}), observe that for each $\pi \in \Sym{W}$ which setwise fixes $S$,
\begin{equation} \label{eqn:part-function-property}
    f^S_{U_\pi^\dagger \rho U_\pi} = f^S_\rho \circ \pi.
\end{equation}
Furthermore, Proposition~\ref{prop:internal-aut-stabilizes} implies that if $\rho$ is a pure state occurring between layers in $C$ given a computational-basis input,
\begin{equation}
    f_\rho \circ \pi = f_\rho \quad\text{for all}~\pi \in \In(C).
\end{equation}

Our indistinguishability argument uses a technical property we call $\ell$-partition symmetry, which broadly states that the reduced state of the circuit on a tuple of wires $\vec w$ after a layer can be determined by the $\ell$-part function of the state before the layer up to a group of symmetries with particularly nice structure. %
Formally, we have the following.
If $(T_i)_i$ is a partition of $\Wd\ell$, we denote by $\prod_i\Sym{T_i}$ the subgroup of $\Sym{\Wd\ell}$ whose permutations do not move elements across partition boundaries. %
We consider $\Aut(C)$ as a subgroup of $\Sym{\Wd\ell}$ through its elementwise action on $\Wd\ell$. %

\begin{definition}
Let $C$ be a $\Gamma_G$-symmetric output quantum circuit and let $L$ represent one of its layers.
$L$ has \defn{$\ell$-partition symmetry} if for every $\vec w \in \Wd\ell$,
there exists a partition $(T_i)_i$ of $\Wd\ell$ such that
$\Stab_{\Aut(C)}(\vec w) \le \prod_i \Sym{T_i}$
and for every $\rho_1,\rho_2 \in \Dens(\Hilb_W)$ representing pure states with entanglement width at most $\ell$,
if $f_{\rho_1} \circ \alpha = f_{\rho_2}$ for some $\alpha \in \prod_i \Sym{T_i}$,
then $\DTr_{\vec w}(L\rho_1L^\dagger) = \DTr_{\vec w}(L\rho_2L^\dagger)$.
$C$ has $\ell$-partition symmetry if the unitaries representing its layers all do.
\end{definition}

In the indistinguishability argument, we use this property to translate an inequality of states on up to $\ell$ wires after a layer to an inequality of $\ell$-part functions up to \textit{every} symmetry $\alpha \in \prod_i \Sym{T_i}$ before the layer.
The remainder of the indistinguishability argument then places pebbles on the support of the orbit $[\vec w']$ of a new tuple of wires $\vec w'$ in order to restore an explicit inequality of states on those wires.
The fact that the inequality of $\ell$-part functions holds up to a large group of symmetries $\alpha$ is used to maintain this inequality and the progress made so far no matter which unitary Duplicator chooses.

For the argument to work, it is important that $\alpha$ comes from a group with the ``partition-symmetric'' structure of $\prod_i \Sym{T_i}$ (\ie, the group contains all symmetries respecting the partition $(T_i)_i$).
Our technique therefore applies to circuits for which we can prove $\ell$-partition symmetry.
In our case, we take $\ell$ equal to the entanglement width $\ew(C)$ of the circuit.

\begin{theorem} \label{thm:partition-symmetry}
Let $G \le \Sym{n}$.
Every $\Gamma_G$-symmetric output quantum circuit $C$ using only 1-qubit and 2-qubit gates has $\ew(C)$-partition symmetry.
\end{theorem}
\begin{proof}
Let $\ew = \ew(C)$.
Suppose $C$ has wires $W$, and observe that $\ew \le \abs{W}$.
Let $\vec w \in \Wd\ew$.
Let $\layer$ be a layer of $C$ represented by $L$, and suppose it consists of the orbits under
$\Aut(C)$ of the applications of some 1-qubit and 2-qubit unitaries $U_1,\dots,U_k$.
We define an equivalence relation $\approx$ on $W$ which expresses that two wires act the same on $\vec w$ in $\layer$.

We define $\approx$ using a sequence of equivalence relations $(\approx_i)_i$, one for each gate $U_i$ (where $i \in [k]$).
Let $w_1,w_2 \in W$.
If $U_i$ is a 1-qubit gate, suppose it applies to the wires $W_{U_i} \subseteq W$ in $\layer$.
Let $w_1 \approx_i w_2$ iff $w_1$ and $w_2$ agree on membership in $W_{U_i}$.

If $U_i$ is a symmetric 2-qubit gate, then for each $w \in \vec w$ and each $v \in W$, let $c_w(v)$ be the number of gates $U_i$ in $\layer$ on the wires $w$ and $v$.
Let $w_1 \approx_i w_2$ iff $c_w(w_1) = c_w(w_2)$ for all $w \in \vec w$.

If $U_i$ is an asymmetric 2-qubit gate, then for each $w \in \vec w$ and $v \in W$, let $c^1_w(v)$ be the number of gates $U_i$ in $\layer$ on wires $(v,w)$ and let $c^2_w(v)$ be the number of such gates on wires $(w,v)$.
Let $w_1 \approx_i w_2$ iff $c^1_w(w_1) = c^1_w(w_2)$ and $c^2_w(w_1) = c^2_w(w_2)$ for all $w \in \vec w$.

Finally, define $\approx$ as follows:
If $w_1 \in \vec w$ or $w_2 \in \vec w$, then $w_1 \approx w_2$ iff $w_1 = w_2$.
Otherwise, $w_1 \approx w_2$ iff $w_1 \approx_i w_2$ for all $i \in [k]$.

Let $(E_i)_i$ be the equivalence classes of $W$ under $\approx$.
Lift $\approx$ to an equivalence relation on $\Wd\ew$ elementwise: if $\vec w_1 = (w_1,\dots,w_\ew), \vec w_2 = (w_1',\dots,w_\ew') \in \Wd\ew$, let $\vec w_1 \approx \vec w_2$ iff $w_i \approx w_i'$ for all $i$.
By the construction of $\approx$, every automorphism $\pi \in \Aut(C)$ which stabilizes $\vec w$ maps each $w \in W$ to $\pi w \approx w$.
Hence, let $(T_i)_i$ be the equivalence classes of $\Wd\ew$ under $\approx$: then $\Stab_{\Aut(C)}(\vec w) \le \prod_i \Sym{T_i}$.

Next, I make the following claim.
\begin{claim} \label{claim:partition-symmetry-core}
Let $\rho_1,\rho_2 \in \Dens(\Hilb_W)$ represent pure states with entanglement width at most $\ew$.
If there is some $\alpha \in \prod_i \Sym{T_i}$ (acting on $\Wd\ew$) such that $f_{\rho_1} \circ \alpha = f_{\rho_2}$, then there is some $\beta \in \prod_i \Sym{E_i}$ (acting on $W$) such that $U_\beta\rho_1U_\beta^\dagger = \rho_2$.
\end{claim}

That is, I claim that if $f_{\rho_1}$ and $f_{\rho_2}$ are related by a suitable permutation of the \textit{set of tuples} $\Wd\ew$, then $\rho_1$ and $\rho_2$ are related by a suitable permutation of the \textit{wires}.
This claim gives us $\ew$-partition symmetry as follows:
Let $\rho_1,\rho_2$ as above and assume $f_{\rho_1} \circ \alpha = f_{\rho_2}$ for some $\alpha \in \prod_i\Sym{T_i}$: then $U_\beta\rho_1U_\beta^\dagger = \rho_2$ for some $\beta \in \prod_i\Sym{E_i}$.
Since all the gates in $\layer$ commute, divide $\layer$ into the set of gates $\layer_1$ which act on at least one of the wires in $\vec w$ and the set of gates $\layer_2$ which do not, and consider the gates in $\layer_2$ to come after those in $\layer_1$.
Represent these sets of gates by unitaries $L_1$ and $L_2$ respectively.  %
By construction, every $\beta \in \prod_i \Sym{E_i}$ does not move any $w \in \vec w$ and is an automorphism of $\layer_1$, so $U_\beta$ commutes with $L_1$, and
by the no-communication theorem, %
$L_2$ cannot affect the partial trace to $\vec w$ of the final state.
Hence we have
\begin{align*}
\DTrTo{\vec w}(L\rho_2L^\dagger) &= \DTrTo{\vec w}(L_2L_1U_\beta\rho_1U_\beta^\dagger L_1^\dagger L_2^\dagger) = \DTrTo{\vec w}(U_\beta L_1\rho_1L_1^\dagger U_\beta^\dagger) = \DTrTo{\vec w}(L_1\rho_1L_1^\dagger) \\
&= \DTrTo{\vec w}(L_2L_1\rho_1L_1^\dagger L_2^\dagger) = \DTrTo{\vec w}(L\rho_1L^\dagger)
\end{align*}
as required.
\begin{innerproof}[{Proof of Claim~\ref{claim:partition-symmetry-core}}]
Let $n = \abs{W}$ and let $\rho_1 = \ketbra{\psi}{\psi}$ and $\rho_2 = \ketbra{\phi}{\phi}$.
Then $\ket\psi$ and $\ket\phi$ have entanglement width at most $\ew$, so let $\ket\psi = \ket{\psi_1} \tprod \cdots \tprod \ket{\psi_r}$ be a decomposition of $\ket\psi$ such that each $\ket{\psi_i}$ is defined over a disjoint set of wires $S_i \subseteq W$ with $\abs{S_i} \le \ew$ which is maximal in the sense that each $\ket{\psi_i}$ is not separable.
The sets $(S_i)_i$ partition $W$.
Let $\ket\phi = \ket{\phi_1} \tprod \cdots \tprod \ket{\phi_{r'}}$ be a maximal decomposition in the same sense.
We argue that a condition analogous to that of Hall's theorem holds.

Suppose that $\vec w_1,\dots,\vec w_k$ are pairwise disjoint $c$-tuples in $\Wd{c}$ for some $c \le \ew$ such that $\vec w_1 \approx \cdots \approx \vec w_k$ and $\DTrTo{\vec w_1}(\rho_1) = \cdots = \DTrTo{\vec w_k}(\rho_1) = \ketbra\xi\xi$ for some $\ket\xi \in (\C^2)^{\tprod c}$, where $\ket\xi$ is not separable.
(Then $\ket\xi$ is a permutation of one of $\ket{\psi_1},\dots,\ket{\psi_r}$.)
I claim that there are at least $k$ pairwise disjoint $c$-tuples $\vec v_i \in \Wd{c}$ in the same equivalence class under $\approx$ such that $\DTrTo{\vec v_i}(\rho_2) = \ketbra\xi\xi$.

We have $c \le \ew$.
Let $d \ge 1$ be the largest integer such that $dc \le \ew$.
If $d \ge k$, then $kc \le \ew$, so there is some $\vec w \in \Wd\ew$ starting with $\vec w_1 \oplus \cdots \oplus \vec w_k$,
where $\oplus$ denotes tuple concatenation. %
Then since $f_{\rho_1} \circ \alpha = f_{\rho_2}$, we have $f_{\rho_1}(\vec w) = f_{\rho_2}(\alpha^{-1} \vec w)$;
hence $\DTrTo{\vec w}(\rho_1) = \DTrTo{\alpha^{-1}\vec w}(\rho_2)$, and so $\alpha^{-1} \vec w$ starts with $\vec v_1 \oplus \cdots \oplus \vec v_k$ where $\vec v_1,\dots,\vec v_k$ are pairwise disjoint $c$-tuples such that $\DTrTo{\vec v_1}(\rho_2) = \cdots = \DTrTo{\vec v_k}(\rho_2) = \ketbra\xi\xi$.
Furthermore, since $\alpha \in \prod_i\Sym{T_i}$, we have $\vec w \approx \alpha^{-1}\vec w$, so $\vec v_i \approx \vec w_i$ for all $i \in [k]$.
So each $\vec v_i$ is in the same equivalence class as $\vec w_1,\dots,\vec w_k$, %
and so at least $k$ such tuples $\vec v_i$ exist.

Otherwise, $d < k$.
Let $\Aut(\ket\xi) := \{\pi \in \Sym{c} : U_\pi\ket\xi = \ket\xi, (\vec w_1)^\pi \approx \vec w_1\}$ be the group of symmetries of $\ket\xi$ with respect to permutations of its $c$ qubits (the \defn{automorphisms} of $\ket\xi$) which preserve the equivalence class under $\approx$ to which $\vec w_1,\dots,\vec w_k$ belong.
Here $(\vec a)^\pi$ refers to the tuple obtained by applying the permutation $\pi$ to the \textit{indices} of $\vec a$: if $\vec a = (w_1,\dots,w_c)$ then $(\vec a)^{\pi} = (w_{\pi^{-1}(1)},\dots,w_{\pi^{-1}(c)})$.
There are exactly
\begin{equation} \label{eqn:partition-symmetry-count}
d! \binom{k}{d} \cdot \abs{\Aut(\ket\xi)}^d \cdot (\ew - dc)! \binom{n-dc}{\ew-dc}
\end{equation}
$\ew$-tuples $\vec w \in \Wd\ew$ which start with $(\vec w_{i_1})^{\pi_1} \concat \cdots \concat (\vec w_{i_d})^{\pi_d}$ for some distinct $\vec w_{i_1},\dots,\vec w_{i_d} \in \{\vec w_1,\dots,\vec w_k\}$ and some automorphisms $\pi_1,\dots,\pi_d \in \Aut(\ket\xi)$ preserving the equivalence class: there are $d!\binom{k}{d}$ ways to choose $\vec w_{i_1},\dots,\vec w_{i_d}$ from $\vec w_1,\dots,\vec w_k$, $\abs{\Aut(\ket\xi)}^d$ ways to choose $\pi_1,\dots,\pi_d$, and $(\ew-dc)! \binom{n-dc}{\ew-dc}$ ways to choose the $\ew-dc$ other elements of $\vec w$.
This formula is exact because $(d+1)c > \ew$, so there are no tuples $\vec{w} \in \Wd\ew$ containing $d+1$ of the $c$-tuples $\vec w_i$.

Since $f_{\rho_1} = f_{\rho_2} \circ \alpha^{-1}$, $\alpha^{-1}$ bijectively maps each of the $\ew$-tuples $\vec w$ to a distinct $\vec v \in \Wd\ew$ which starts with $\vec v_1 \oplus \cdots \oplus \vec v_d$, where $\vec v_1,\dots,\vec v_d \in \Wd{c}$ are pairwise disjoint $c$-tuples with $\DTrTo{\vec v_1}(\rho_2) = \cdots = \DTrTo{\vec v_d}(\rho_2) = \ketbra\xi\xi$.
As before, since $\alpha \in \prod_i \Sym{T_i}$, each $\vec v_i$ is in the same equivalence class under $\approx$ as $\vec w_1,\dots,\vec w_k$.

Now since $\ket\xi$ is a non-separable pure state, if $\vec u_1$ and $\vec u_2$ are two such $c$-tuples in $\Wd{c}$ both in the same equivalence class as $\vec w_1,\dots,\vec w_k$ with $\DTrTo{\vec u_1}(\rho_2) = \DTrTo{\vec u_2}(\rho_2) = \ketbra\xi\xi$, then either $\vec u_1$ and $\vec u_2$ are disjoint or $\vec u_1 = (\vec u_2)^\pi$ for some automorphism $\pi \in \Aut(\ket\xi)$.
Indeed, suppose $\vec u_1$ and $\vec u_2$ are not disjoint, and let $S'$ be the set of wires in both $\vec u_1$ and $\vec u_2$.
If $S'$ does not contain all the wires in $\vec u_1$ and $\vec u_2$ (\ie, $\vec u_1$ and $\vec u_2$ do not fully overlap), then since $\DTrTo{\vec u_1}(\rho_2)$ and $\DTrTo{\vec u_2}(\rho_2)$ are pure states, by Lemma~\ref{lem:app-pure-overlap}, $\TrTo{S'}(\rho_2)$ is a pure state,
which (by Lemmas~\ref{lem:app-pure-induces-partition} and \ref{lem:app-pure-product-implies-pure}) violates the non-separability of $\ket\xi$.
Hence $\vec u_1$ and $\vec u_2$ fully overlap, so since $\DTrTo{\vec u_1}(\rho_2) = \DTrTo{\vec u_2}(\rho_2) = \ketbra\xi\xi$ and $\vec u_1 \approx \vec u_2 \approx \vec w_1$, we must have $\vec u_1 = (\vec u_2)^\pi$ for some $\pi \in \Aut(\ket\xi)$.

Hence each pair of the $c$-tuples $\vec v_i$ are either disjoint or are related by an automorphism.
Thus, if there are only at most $k-1$ disjoint $c$-tuples $\vec v_i \in \Wd{c}$ such that $\DTrTo{\vec v_i}(\rho_2) = \ketbra\xi\xi$, then by the same calculation as above, there are only at most
\begin{equation}
  d! \binom{k-1}{d} \cdot \abs{\Aut(\ket\xi)}^d \cdot (\ew - dc)! \binom{n - dc}{\ew - dc}
\end{equation}
possible $\ew$-tuples $\vec v$ as defined above.
Since the $\vec v$ are mapped bijectively by $\alpha$ to the $\ew$-tuples $\vec w$, whose count is given by (\ref{eqn:partition-symmetry-count}), this is not enough!
So there are at least $k$ such disjoint $c$-tuples $\vec v_i$ in the same equivalence class as $\vec w_1,\dots,\vec w_k$.

By symmetry, it also holds that if $\vec{v}_1,\dots,\vec{v}_k$ is a set of disjoint $c$-tuples in $\Wd{c}$ for some $c \le \ew$ such that $\vec v_1 \approx \cdots \approx \vec v_k$ and $\DTrTo{\vec{v}_1}(\rho_2) = \cdots = \DTrTo{\vec{v}_k}(\rho_2) = \ketbra\xi\xi$ for some non-separable pure state $\ket\xi$ (\ie, a permutation of one of $\ket{\phi_1},\dots,\ket{\phi_{r'}}$), then there are at least $k$ disjoint $c$-tuples $\vec{w}_i \in \Wd{c}$ in the same equivalence class under $\approx$ as $\vec v_1,\dots,\vec v_k$ such that $\DTrTo{\vec w_i}(\rho_1) = \ketbra\xi\xi$.

Therefore there is a bijection $g : [r] \to [r']$ (and so $r = r'$) between $(\ket{\psi_i})_{i=1}^r$ and $(\ket{\phi_i})_{i=1}^{r'}$ with the following property:
For every $i \in [r]$, let $\vec w_i$ be an ordering of $S_i$, the set of wires underlying $\ket{\psi_i}$.
There is an ordering $\vec v_{g(i)}$ of the set of wires underlying $\ket{\phi_{g(i)}}$ such that:
\begin{enumerate}[label=(\alph*)]
    \item \label{lbl:partition-symmetry-prop1} The states $\ket{\psi_i}$ and $\ket{\phi_{g(i)}}$ are equal up to a reordering of the wires of $\ket{\psi_i}$ to $\vec w_i$ and of $\ket{\phi_{g(i)}}$ to $\vec v_{g(i)}$; \ie, $\DTrTo{\vec w_i}(\rho_1) = \DTrTo{\vec v_{g(i)}}(\rho_2)$; and
    \item \label{lbl:partition-symmetry-prop2} $\vec w_i \approx \vec v_{g(i)}$.
\end{enumerate}
Let $\vec w = \vec w_1 \oplus \cdots \oplus \vec w_r$ and $\vec v = \vec v_{g(1)} \oplus \cdots \oplus \vec v_{g(r)}$; both are orderings of all of $W$.
Hence $\vec v = \pi \vec w$ for some $\pi \in \Sym{W}$.
By property \ref{lbl:partition-symmetry-prop2}, $\vec w \approx \vec v$.
Let $\vec w = (w_1,\dots,w_n)$ and $\vec v = (v_1,\dots,v_n)$: we have $v_j = \pi w_j$ and $v_j \approx w_j$ for each $j \in [n]$, so since $w_1,\dots,w_n$ and $v_1,\dots,v_n$ enumerate $W$,
$\pi$ maps members of each equivalence class of $W$ under $\approx$ only to members of the same equivalence class.
Hence in fact $\pi \in \prod_i \Sym{E_i}$.
Finally, by property \ref{lbl:partition-symmetry-prop1},
\begin{equation}
    \DTrTo{\vec w}(\rho_1) = \DTrTo{\vec v}(\rho_2) = \DTrTo{\pi \vec w}(\rho_2) = \DTrTo{\vec w}(U_\pi^\dagger \rho_2 U_\pi)
\end{equation}
where the last equality is by (\ref{eqn:dtr-property}).
So $\rho_1 = U_\pi^\dagger \rho_2 U_\pi$.
\qedhere
\end{innerproof}
\end{proof}

\section{Indistinguishability proof}\label{sec:indistinguishability}

We now prove our indistinguishability theorem: if a symmetric quantum circuit $C$ has $\ell$-partition symmetry and supports of size at most $k$, then it cannot distinguish inputs on which Duplicator wins the partial trace-based quantum pebble game with $2k$ pebbles.
This is a quantum generalization of Theorem 14 of \cite{Dawar2024} (Theorem 4.8 in the full version) and our proof strategy largely generalizes theirs.

Given a circuit $C$ with $\maxsupp_\ell(C) \le k$, the following lemma shows we can assign a support $\supp([\vec w])$ to each $[\vec w] \in \Wd\ell/\In(C)$ in a manner respecting the automorphisms of the circuit.
Recall that $G$ acts on $\Wd\ell/\In(C)$ through its embedding into $\Aut(C)/\In(C)$ (Proposition~\ref{prop:embedding}).

\begin{lemma}
Suppose $P \subseteq \Peb_G$ supports $[\vec w] \in \Wd\ell/\In(C)$.
Then for all $\sigma \in G$, $\sigma P \sigma^{-1} = \{\sigma \Gamma_p \sigma^{-1} : \Gamma_p \in P\}$ supports $\sigma [\vec w]$.
\end{lemma}
\begin{proof}
Observe that $\inter (\sigma P \sigma^{-1}) = \sigma (\inter P)\sigma^{-1}$.
That $P$ supports $[\vec w]$ means $\inter P \le \Stab_G([\vec w])$, so $\sigma(\inter P)\sigma^{-1} \le \Stab_G(\sigma [\vec w])$.
\end{proof}

Therefore, to each $[\vec w] \in \Wd\ell/\In(C)$ we assign a support $\supp([\vec w]) \subseteq \Peb_G$ such that $\supp(\sigma[\vec w]) = \sigma\supp([\vec w])\,\sigma^{-1}$.
If $\maxsupp_\ell(C) \le k$, we may take $\abs{\supp([\vec w])} \le k$.

We are now ready to prove the indistinguishability theorem.

\begin{theorem}
\label{thm:indistinguishability}
Let $G \le \Sym{n}$ with distinct stabilizers.
Let $C$ be a $\Gamma_G$-symmetric output quantum $\tau$-circuit with at least $\ell$ wires which has $\ell$-partition symmetry, has $\ew(C) \le \ell$, and has $\maxsupp_\ell(C) \le k$.
Let $\str{A},\str{B} \in \fin_n[\tau]$.
If Duplicator wins the $(\Gamma_G,\Peb_G,2k)$-$\ell$-PT quantum pebble game on $\ket{\str{A}}$ and $\ket{\str{B}}$, then $C[\ket{\str{A}}] = C[\ket{\str{B}}]$.
\end{theorem}
\begin{proof}
We prove the contrapositive: assume $C[\ket{\str{A}}] \ne C[\ket{\str{B}}]$.
We construct a winning Spoiler strategy for the game on $\ket{\str{A}}$ and $\ket{\str{B}}$.
Suppose $C$ has wires $W = X \sqcup A$, where $X$ are the input wires and $A$ are the ancillas.
Let the layers of $C$ be represented by unitaries $L_1,\dots,L_m$, let $L_{\le i} = L_iL_{i-1}\cdots L_1$ for $i \in [m]$ and let $L_{\le 0} = \iden$, and for each $0 \le i \le m$, let
$\rho_i^{\str{A}} = L_{\le i}(\ketbra{\str{A}}{\str{A}} \tprod \ketbra00^{\tprod A})L_{\le i}^\dagger$
and
$\rho_i^{\str{B}} = L_{\le i}(\ketbra{\str{B}}{\str{B}} \tprod \ketbra00^{\tprod A})L_{\le i}^\dagger$.

For each $i$ from $m$ to $0$, Spoiler plays to maintain the invariant that there is some $\vec w_i \in \Wd{c_i}$ for some $c_i \le \ell$ such that $[\vec w_i]$ is supported by the current set of pebbles $P$ and for every automorphism $\pi$ induced by the game's current unitary $\sigma \in G$ (again identifying $G$ with $\Gamma_G$ by $\sigma \mapsto U_\sigma$), we have $\DTrTo{\vec w_i}(\rho_i^\str{A}) \ne \DTrTo{\vec w_i}(U_\pi^\dagger \rho_i^{\str{B}} U_\pi)$.

Then for $i = 0$, let $S$ be the set of wires in $\vec w_0$:
we have $\TrTo{S}(\rho_0^{\str{A}}) \ne \TrTo{S}(U_\pi^\dagger \rho_0^{\str{B}} U_\pi)$ for all $\pi$ induced by $\sigma$, where $\rho_0^{\str{A}} = \ketbra{\str{A}}{\str{A}} \tprod \ketbra00^{\tprod A}$ and $\rho_0^{\str{B}} = \ketbra{\str{B}}{\str{B}} \tprod \ketbra00^{\tprod A}$.
Since each automorphism permutes the input wires and ancillas separately, if $S$ contains only ancillas, then $\TrTo{S}(\rho_0^{\str{A}}) = \TrTo{S}(U_\pi^\dagger\rho_0^{\str{B}}U_\pi) = \ketbra00^{\tprod S}$; but this is not the case, so $S$ contains some input wires.
Let $S_X = S \cap X$: then $1 \le \abs{S_X} \le \ell$.
Furthermore, since $P$ supports $[\vec w_0]$ and the input wires are not moved by any internal automorphism, %
$S_X \subseteq Q^\tau_P$, \ie, $\inter P$ pointwise stabilizes $S_X$.
We have $\TrTo{S_X}(\ketbra{\str{A}}{\str{A}}) \ne \TrTo{S_X}(U_\sigma^\dagger\ketbra{\str{B}}{\str{B}}U_\sigma)$ (recall that $U_\sigma$ acts on $\Hilb^\tau_n$).
This is a violation of $\ell$-PT partial isomorphism from $\ket{\str{A}}$ to $\ket{\str{B}}$, so Spoiler wins.

For the base case $i = m$, recall that the output wire $w_\out$ of $C$ is not moved by any automorphism, so let $c_m = 1$ and $\vec w_m = (w_\out)$.
Then $[\vec w_m] = \{\vec w_m\}$ is supported by the empty set of pebbles.
$C[\ket{\str{A}}] \ne C[\ket{\str{B}}]$ gives us $\TrTo{\vec w_m}(\rho_m^{\str{A}}) \ne \TrTo{\vec w_m}(U_\pi^\dagger \rho_m^{\str{B}} U_\pi)$ for all $\pi \in \Aut(C)$.

The inductive step follows from the following claim, analogous to Claim 4.9 in the full version of \cite{Dawar2024}.
\begin{claim}
Suppose we are at layer $0 \le i < m$ and let $L = L_{i+1}$, $\rho_\str{A} = \rho_i^{\str{A}}$, and $\rho_\str{B} = \rho_i^{\str{B}}$.
Suppose the current pebbles include $P \subseteq \Peb_G$ with $\abs{P} \le k$, the current unitary is $\sigma \in G$, and we have $\vec w = \vec w_{i+1} \in \Wd{c}$ for some $c = c_{i+1} \le \ell$ such that $P$ supports $[\vec w]$ and for every automorphism $\pi$ induced by $\sigma$,
\begin{equation}
    \DTrTo{\vec w}(L\rho_\str{A}L^\dagger) \ne \DTrTo{\vec w}(U_\pi^\dagger L\rho_\str{B}L^\dagger U_\pi).
\end{equation}
Spoiler can play such that after at most $k$ rounds, we are in a position where the pebbles include $P' \subseteq \Peb_G$ with $\abs{P'} \le k$, the unitary is $\sigma' \in G$, and we have some $\vec w' \in \Wd\ell$ such that $P'$ supports $[\vec w']$ and for every automorphism $\pi'$ induced by $\sigma'$,
\begin{equation}
    \DTrTo{\vec w'}(\rho_\str{A}) \ne \DTrTo{\vec w'}(U_{\pi'}^\dagger \rho_\str{B} U_{\pi'}).
\end{equation}
\end{claim}
\begin{proof}
Since $C$ has $\ell$-partition symmetry, there is a partition $(T_i)_i$ of $\Wd\ell$ such that $\Stab_{\Aut(C)}(\vec w) \le \prod_i \Sym{T_i}$ and $f_{\rho_\str{A}} \circ \alpha \ne f_{U_\pi^\dagger\rho_\str{B}U_\pi}$ for all $\alpha \in \prod_i \Sym{T_i}$ and all $\pi$ induced by $\sigma$.
Then $\chi(\inter P) \cap \Stab_{\Aut(C)}(\vec w) \le \prod_i \Sym{T_i}$, so the orbits of $\Wd\ell$ under $\chi(\inter P) \cap \Stab_{\Aut(C)}(\vec w)$ form a refinement of $(T_i)$.
Hence there is some such orbit $O$ such that
\begin{equation}
    f^O_{\rho_\str{A}} \circ \alpha \ne f^O_{U_{\pi}^\dagger \rho_\str{B} U_\pi}
\end{equation}
for all $\alpha \in \Sym{O}$ and all $\pi$ induced by $\sigma$.

We now place pebbles.
Suppose that in the $j$th round, the pebbles include $P_j = P \cup \{\Gamma_{p_1},\dots,\Gamma_{p_j}\}$ and the current unitary is $\sigma_j$.
Spoiler plays to maintain the following invariant:
there is some $\vec w_j' \in \Wd\ell$ such that $\Gamma_{p_1},\dots,\Gamma_{p_j} \in \supp([\vec w_j'])$ and for its orbit $O_j = \Orb_{\chi(\inter P_j) \cap \Stab_{\Aut(C)}(\vec w)}(\vec w'_j)$,
\begin{equation}
f^{O_j}_{\rho_\str{A}} \circ \alpha \ne f^{O_j}_{U_{\pi_j}^\dagger \rho_\str{B} U_{\pi_j}}
\end{equation}
for all $\alpha \in \Sym{O_j}$ and all $\pi_j$ induced by $\sigma_j$.
Since $\abs*{\supp([\vec w_j'])} \le k$, after at most $k$ rounds, we can take $\alpha = \mathrm{id}$ and apply $\vec w_j'$ to get that $\DTrTo{\vec w_j'}(\rho_\str{A}) \ne \DTrTo{\vec w_j'}(U_{\pi_j}^\dagger \rho_\str{B} U_{\pi_j})$ for all $\pi_j$ induced by $\sigma_j$ with $[\vec w_j']$ supported by $\{\Gamma_{p_1},\dots,\Gamma_{p_j}\}$, which proves the claim.

The case $j = 0$ is given above.
In round $j \ge 1$, Spoiler first picks up some pebble not in the support of $\vec w$ or $\vec w'_{j-1}$ if necessary.
From the invariant,
\begin{equation} \label{eqn:indist-intermediate-1}
    f^{O_{j-1}}_{\rho_\str{A}} \circ \alpha \ne f^{O_{j-1}}_{U_{\pi_{j-1}}^\dagger \rho_\str{B} U_{\pi_{j-1}}}
\end{equation}
for all $\alpha \in \Sym{O_{j-1}}$ and all $\pi_{j-1}$ induced by $\sigma_{j-1}$.
Duplicator chooses some $\sigma \in \inter P_{j-1}$ and sets $\sigma_j := \sigma_{j-1}\sigma$.
Observe that every automorphism $\pi_j$ induced by $\sigma_j$ is of the form $\pi_{j-1}\pi$, where $\pi_{j-1}$ and $\pi$ are induced by $\sigma_{j-1}$ and $\sigma$ respectively.
By equation (\ref{eqn:part-function-property}),
\begin{equation}
    f_{U_{\pi_j}^\dagger \rho_\str{B} U_{\pi_j}} = f_{U_{\pi_{j-1}}^\dagger \rho_\str{B} U_{\pi_{j-1}}} \circ \pi.
\end{equation}
Since $P$ supports $[\vec w]$, $\sigma$ stabilizes $[\vec w]$.
Lemma~\ref{lem:g-to-aut-stabilizes} states that $\sigma$ induces some automorphism $\pi'$ stabilizing $\vec w$.
By Proposition~\ref{prop:embedding}, $\pi'$ differs from $\pi$ only by some internal automorphism, so by equation (\ref{eqn:part-function-property}), in fact
\begin{equation} \label{eqn:indist-intermediate-2}
    f_{U_{\pi_j}^\dagger \rho_\str{B} U_{\pi_j}} = f_{U_{\pi_{j-1}}^\dagger \rho_\str{B} U_{\pi_{j-1}}} \circ \pi'.
\end{equation}
Now $\pi' \in \chi(\inter P_{j-1}) \cap \Stab_{\Aut(C)}(\vec w)$ and $O_{j-1}$ is the orbit of $\vec w'_{j-1}$ under this group, so we may restrict both sides of (\ref{eqn:indist-intermediate-2}) to $O_{j-1}$:
\begin{equation}
    f_{U_{\pi_j}^\dagger \rho_\str{B} U_{\pi_j}}^{O_{j-1}} = f_{U_{\pi_{j-1}}^\dagger \rho_\str{B} U_{\pi_{j-1}}}^{O_{j-1}} \circ \pi'
\end{equation}
Combining with (\ref{eqn:indist-intermediate-1}) and observing that $\pi' \in \Sym{O_{j-1}}$, we obtain
\begin{equation} \label{eqn:indist-defused}
    f^{O_{j-1}}_{\rho_\str{A}} \circ \alpha \ne f^{O_{j-1}}_{U_{\pi_{j}}^\dagger \rho_\str{B} U_{\pi_{j}}}
\end{equation}
for all $\alpha \in \Sym{O_{j-1}}$ and all $\pi_j$ induced by $\sigma_j$.

Having defused Duplicator's choice of unitary, we now show how Spoiler can make progress.
Let $\Gamma_p$ be some pebble in $\supp([\vec w_{j-1}'])$ which is not yet in $P_{j-1}$.
For each pebble $\Gamma_q \in \Peb_G$ which is conjugate to $\Gamma_p$ by some $\pi \in \chi(\inter P_{j-1}) \cap \Stab_{\Aut(C)}(\vec w)$, let
\begin{equation}
W_q := \{\pi \vec w_{j-1}' : \pi \in \Stab_{\Aut(C)}(\vec w) \text{ is induced by some } \sigma \in \inter P_{j-1} \text{ such that } \sigma \Gamma_p \sigma^{-1} = \Gamma_q \}.
\end{equation}
We make a series of claims about these sets $(W_q)_q$:
\begin{enumerate}[label=(\roman*)]
    \item Every $\vec v \in O_{j-1}$ appears in the same number of sets $W_q$.
    \item There is some $\Gamma_q$ such that $f^{W_q}_{\rho_\str{A}} \circ \alpha \ne f^{W_q}_{U_{\pi_j}^\dagger \rho_\str{B} U_{\pi_j}}$ for all $\alpha \in \Sym{W_q}$ and all $\pi_j$ induced by $\sigma_j$.
    \item For every $\Gamma_q$, we have $W_q = \Orb_{\chi(\Gamma_q \cap \inter P_{j-1}) \cap \Stab_{\Aut(C)}(\vec w)}(\vec v)$ for $\vec v \in W_q$.
\end{enumerate}

\textit{For (i):}
Let $\vec v \in O_{j-1}$: then $\vec v = \pi \vec w_{j-1}'$ for some $\pi \in \Stab_{\Aut(C)}(\vec w)$ induced by some $\sigma \in \inter P_{j-1}$.
The set of $\Gamma_q$ such that $\vec v \in W_q$ is given by $\{\sigma'\Gamma_p\sigma'^{-1} : \pi' \vec w_{j-1}' = \vec v \text{ for some } \pi' \in \Stab_{\Aut(C)}(\vec w) \text{ induced by } \sigma' \in \inter P_{j-1}\}$.
There is a bijection to this set from the fixed set $\{\sigma' \Gamma_p \sigma'^{-1} : \pi' \vec w_{j-1}' = \vec w_{j-1}' \text{ for some } \pi' \in \Stab_{\Aut(C)}(\vec w) \text{ induced by } \sigma' \in \inter P_{j-1}\}$ given by conjugating by $\sigma$.

\textit{For (ii):}
Suppose there is no such $\Gamma_q$.
Then for each $\Gamma_q$ there is some $\alpha \in \Sym{W_q}$ such that $f^{W_q}_{\rho_\str{A}} \circ \alpha = f^{W_q}_{U_{\pi_j}^\dagger \rho_\str{B} U_{\pi_j}}$ for some $\pi_j$ induced by $\sigma_j$.
Fix a particular automorphism $\pi$ induced by $\sigma_j$: then $\pi\pi_j^{-1} \in \In(C)$, so applying (\ref{eqn:part-function-property}) and Lemma~\ref{lem:g-to-aut-stabilizes} gives
\begin{equation}
    f_{U_{\pi_j}^\dagger \rho_\str{B} U_{\pi_j}} = f_{U_\pi^\dagger \rho_\str{B} U_\pi}.
\end{equation}
Restricting to the domain $W_q$, for each $\Gamma_q$, we have $f_{\rho_\str{A}}^{W_q} \circ \alpha = f^{W_q}_{U_\pi^\dagger \rho_\str{B} U_\pi}$ for some $\alpha \in \Sym{W_q}$.
That is, the multisets given by the images of $f^{W_q}_{\rho_\str{A}}$ and $f^{W_q}_{U_{\pi}^\dagger\rho_\str{B}U_{\pi}}$ are equal.
By claim (i), taking the union of these multisets for all $\Gamma_q$ gives a multiset in which the image of each $\vec v \in O_{j-1}$ appears the same number of times (say $t$ times).
That is, the resulting multiset consists of $t$ copies of the images of $f^{O_{j-1}}_{\rho_\str{A}}$ and $f^{O_{j-1}}_{U_{\pi}^\dagger\rho_\str{B}U_{\pi}}$, so these images are equal as multisets.
Therefore, $f^{O_{j-1}}_{\rho_\str{A}} \circ \alpha = f^{O_{j-1}}_{U_\pi^\dagger\rho_\str{B}U_\pi}$ for some $\alpha \in \Sym{O_{j-1}}$, contradicting (\ref{eqn:indist-defused}).

\textit{For (iii):}
For brevity, let $Q = \Gamma_q \cap \inter P_{j-1}$.
Let $\vec v \in W_q$: there is some $\pi \in \Stab_{\Aut(C)}(\vec w)$ induced by some $\sigma \in \inter P_{j-1}$ such that $\vec v = \pi \vec w_{j-1}'$ and $\sigma \Gamma_p \sigma^{-1} = \Gamma_q$.
For every $\pi' \in \chi(Q) \cap \Stab_{\Aut(C)}(\vec w)$ induced by $\sigma' \in Q$, we have $(\sigma'\sigma)\Gamma_p(\sigma'\sigma)^{-1} = \sigma'\Gamma_q\sigma'^{-1} = \Gamma_q$ since $\sigma' \in \Gamma_q$, so since $\sigma'\sigma$ induces $\pi'\pi$, we have $\pi'\vec v = \pi'\pi \vec w_{j-1}' \in W_q$.
Hence $\Orb_{\chi(Q) \cap \Stab_{\Aut(C)}(\vec w)}(\vec v) \subseteq W_q$.
Conversely, for all $\pi' \in \Stab_{\Aut(C)}(\vec w)$ induced by $\sigma' \in \inter P_{j-1}$ such that $\sigma' \Gamma_p \sigma'^{-1} = \Gamma_q$ (i.e., for every $\pi' \vec w_{j-1}' \in W_q$), we have $(\sigma'\sigma^{-1})\Gamma_q(\sigma'\sigma^{-1})^{-1} = \sigma'\Gamma_p\sigma'^{-1} = \Gamma_q$.
But, since $G$ has distinct stabilizers, every pebble is a self-normalizing subgroup; thus we have $\sigma'\sigma^{-1} \in \Gamma_q$ and so $\sigma'\sigma^{-1} \in Q$.
Hence $\pi'\pi^{-1} \in \chi(Q) \cap \Stab_{\Aut(C)}(\vec w)$.
Furthermore, $\pi'\pi^{-1} \vec v = \pi' \vec w_{j-1}'$, so $\pi'\vec w_{j-1}' \in \Orb_{\chi(Q) \cap \Stab_{\Aut(C)}(\vec w)}(\vec v)$.
Hence $W_q \subseteq \Orb_{\chi(Q) \cap \Stab_{\Aut(C)}(\vec w)}(\vec v)$ as well.

Thus, by claim (ii), Spoiler places a pebble $\Gamma_q$ such that $f^{W_q}_{\rho_\str{A}} \circ \alpha \ne f^{W_q}_{U_{\pi_j}^\dagger \rho_\str{B} U_{\pi_j}}$ for all $\alpha \in \Sym{W_q}$ and all $\pi_j$ induced by $\sigma_j$.
Let $P_j := P_{j-1} \cup \{\Gamma_q\}$.
Choose any $\vec w_j' \in W_q$: we have $\vec w_j' = \pi \vec w_{j-1}'$ (so $[\vec w_j'] = \sigma [\vec w_{j-1}']$) for some $\pi$ induced by $\sigma \in \inter P_{j-1}$ with $\Gamma_q = \sigma \Gamma_p \sigma^{-1}$, so by our assumption on the assignment of supports, $\Gamma_{p_1}, \dots, \Gamma_{p_{j-1}}, \Gamma_q \in \supp([\vec w_j'])$.
By claim (iii), let $O_j := W_q = \Orb_{\chi(\inter P_j) \cap \Stab_{\Aut(C)}(\vec w)}(\vec w_j')$.
This preserves the invariant.
\end{proof}
Repeatedly applying this claim gives a Spoiler strategy that wins in at most $km$ rounds.
\end{proof}

\section{CFI application and parameter calculation}
\label{sec:cfi}

In this section, we define the graph property~$\PAltCFI$ that is hard for moderate-entanglement symmetric quantum circuits to solve, and prove the corresponding lower bound (Theorem~\ref{thm:main}).
We will in fact prove the following stronger result concerning quantum circuits which are symmetric under the action of the alternating group $\Alt{n}$, from which Theorem~\ref{thm:main} immediately follows.

\begin{theorem}\label{thm:mainAlt}
Let $f:\Z_+\to\Z_+$ be any increasing function with $f(n)=\omega(\log n)$. There is no family of $\Gamma_{\Alt{n}}$-symmetric output quantum $\taugraph$-circuits $(C_n)_n$ computing $\PAltCFI$ which has size $s(C_n) = 2^{o(f(n))}$ and entanglement width $\eta(C_n) = O(n/f(n))$.
\end{theorem}

\subsection{The CFI and Alt-CFI graphs}

For a given logic $\logic{L}$ and structures ${\str{A}}$ and ${\str{B}}$ over the same
vocabulary, write ${\str{A}}\equiv_{\logic{L}}{\str{B}}$ to mean that ${\str{A}}$ and ${\str{B}}$ agree on every sentence in $\logic{L}$.
We follow the notation of Immerman and Lander \cite{Immerman1990}.

\begin{definition}
Let $\logic{L}_k$ denote the fragment of first-order logic consisting of formulas in
which the only variables that appear are from the finite set
$\{x_1,\ldots,x_k\}$.
\end{definition}

\begin{definition}
Let $\logic{C}_k$ denote the extension of $\logic{L}_k$ with the following syntax rule:
if $n\in\Z_{\geq1}$, $x\in\{x_1,\ldots,x_k\}$ is a variable, and
$\phi$ is a formula, then $\exists^{\geq n}x\,\phi$ is a formula.
\end{definition}

The semantics are as follows: if $\str{A}$ is a structure with domain $A$,
$\phi(x)$ is a formula, and $n\in\Z_{\geq1}$, then
$\str{A}\models\exists^{\geq n}x\,\phi(x)$ if and only if there are at least
$n$ distinct elements $a\in A$ such that $\str{A}\models\phi[a]$.

Cai, F\"urer, and Immerman~\cite{Cai1992} constructed a remarkable
counterexample showing that $\FPC$, even on $\taugraph$, does
not capture all polynomial-time properties. We present their results
but omit the details of the construction. The interested reader may
refer to~\cite{Cai1992}.

\begin{theorem}[\cite{Cai1992}]
\label{thm:classical-cfi}
There exist families of undirected graphs $(G_n)_n$ and $(H_n)_n$,
$n\in\Z_{\geq 1}$, called the \defn{CFI graphs}, with the following
properties for each $n$:
\begin{enumerate}
  \item $G_n$ and $H_n$ have the same number $v=O(n)$ of vertices,
        with $v\geq n+1$.
  \item $G_n\equiv_{\logic{C}_n}H_n$ when interpreted as
        $\taugraph$-structures.
  \item $G_n$ is not isomorphic to $H_n$.
\end{enumerate}
Furthermore, there is a polynomial-time algorithm which outputs
``yes'' on every $G_n$ and ``no'' on every $H_n$.
\end{theorem}

The ``furthermore'' part of Theorem~\ref{thm:classical-cfi} implies
that there is a polynomial-time $\taugraph$-property, the
\defn{CFI property} $\PCFI$, such that $G_n\in \PCFI$
but $H_n\notin \PCFI$ for each $n$.

\begin{definition}
Let $A'\subseteq A$ and $B'\subseteq B$.
A bijection $f:A'\to B'$ is a \defn{partial isomorphism} from $\str{A}$ to $\str{B}$
if for every relation symbol $R_i^{r_i}\in\tau$ and every
$a_1,\ldots,a_{r_i}\in A'$, we have
$(a_1,\ldots,a_{r_i})\in R_i^{\str{A}}$ if and only if
$(f(a_1),\ldots,f(a_{r_i}))\in R_i^{\str{B}}$.
\end{definition}

Let ${\str{A}}$ and ${\str{B}}$ be finite $\tau$-structures of the same size $n$, and
suppose without loss of generality that they have the domain $[n]$.
Let $G$ be a group acting on $[n]$. Without loss of generality, we
may take $G\leq\Sym{n}$.

\begin{definition}[\cite{Dawar2024}]
\label{def:classical-game}
The \defn{$k$-pebble $G$-bijection game} on ${\str{A}}$ and ${\str{B}}$ is played
using two sets of $k$ pebbles, $a_1,\ldots,a_k$ and
$b_1,\ldots,b_k$. Initially, no pebbles are placed. Each round
proceeds as follows:
\begin{enumerate}
  \item Spoiler picks up a pair of pebbles $(a_i,b_i)$.
  \item Duplicator chooses some $\sigma\in G$ such that for every
        pair of pebbles $(a_j,b_j)$ which are placed on the
        structures, we have $\sigma(a_j)=b_j$.
  \item Spoiler places $a_i$ on some unpebbled element $a$ of ${\str{A}}$
        and places $b_i$ on $\sigma(a_i)$.
\end{enumerate}
Spoiler wins if the mapping given by the pebbles is not a partial
isomorphism. Otherwise, play continues, and Duplicator wins if it has
a strategy to play forever.
\end{definition}

We abbreviate ``$k$-pebble $G$-bijection game'' to
``$(G,k)$-pebble game''. To recover Hella's bijection game, take
$G=\Sym{n}$.

\begin{corollary}\label{cor:counting-game}
Duplicator wins the $k$-pebble bijection game on ${\str{A}}$ and ${\str{B}}$
if and only if ${\str{A}}\equiv_{\logic{C}_k}{\str{B}}$.
\end{corollary}

\begin{theorem}[\cite{Dawar2015}]
\label{thm:sym-cfi}
Let $(G_n)_n$ and $(H_n)_n$ be the families of CFI graphs of
Theorem~\ref{thm:classical-cfi}, interpreted as
$\taugraph$-structures, and suppose $G_n$ and $H_n$
have the same set of vertices $V$. Then Duplicator wins the
$(\Sym{V},n)$-pebble game on $G_n$ and $H_n$.
\end{theorem}

\begin{proof}
$G_n\equiv_{\logic{C}_n}H_n$ by Theorem~\ref{thm:classical-cfi}, so
Duplicator wins the $n$-pebble bijection game on $G_n$ and $H_n$ by Corollary~\ref{cor:counting-game}.
Thus Duplicator wins the $(\Sym{V},n)$-pebble game.
\end{proof}

It is convenient to instead have families of structures on which
Duplicator wins the $(\Alt{n},k)$-pebble game, where
$\Alt{n}\leq\Sym{n}$ is the alternating
group consisting of all even permutations of $[n]$~\cite{Dixon1996}.
The following theorem provides a way to translate Duplicator
strategies for the $(\Sym{n},k)$-pebble game on graphs
to the $(\Alt{n},k)$-pebble game.

\begin{theorem}
\label{thm:alt-padding}
Let $G$ and $H$ be nonisomorphic graphs on the same set of vertices
$V$, interpreted as $\taugraph$-structures, such that
Duplicator wins the $(\Sym{V},k)$-pebble game on $G$
and $H$. There are nonisomorphic graphs $G'$ and $H'$ on vertices
$V'$ with $\abs{V'}=\abs{V}+k+2$ such that Duplicator wins the
$(\Alt{V'},k)$-pebble game on $G'$ and $H'$.
\end{theorem}

\begin{proof}
Add $k+2$ isolated vertices to each of $G$ and $H$ to form $G'$
and $H'$. Let $I$ be the set of isolated vertices added. We have
$V'=V\cup I$. In the $(\Alt{V'},k)$-pebble game
on $G'$ and $H'$, Duplicator will play only bijections that map
vertices in $V$ to vertices in $V$ and vertices in $I$ to vertices
in $I$, so no pair of pebbles $(a_i,b_i)$ will ever map a vertex
in $V$ to a vertex in $I$ or vice versa.

Our strategy is as follows. Suppose that in the
$(\Sym{V},k)$-pebble game, ignoring the vertices $I$,
Duplicator would play $\sigma\in\Sym{V}$. Let
$\pi\in\Sym{I}$ be a bijection on $I$ respecting the
pebbles placed on $I$, \ie, such that for every pair of pebbles
$(a_j,b_j)$ placed on $I$, $\pi(a_j)=b_j$. Since there are only
$k$ pebbles and $\abs{I}=k+2$, there are at least two distinct isolated
vertices $i_1,i_2\in I$ which are unpebbled in $G'$. So
$\pi(i_1\ i_2)$ also respects the pebbles on $I$. Thus, if
$\sigma\pi$, considered by abuse of notation as an element of
$\Sym{V'}$, is an even permutation, Duplicator
plays $\sigma\pi$. If $\sigma\pi$ is odd, then it plays
$\sigma\pi(i_1\ i_2)$, which is even. This strategy simulates the
$(\Sym{V},k)$-pebble game strategy on the reduction
to $V$ and wins since the pebbles on $I$ cannot cause a partial
isomorphism violation.
\end{proof}

We usually have $k\leq \abs{V}$ since Spoiler always wins if there are
more pebbles than domain elements and the structures are
nonisomorphic. Then $\abs{V'}\leq 2\abs{V}+2$ and so
Theorem~\ref{thm:alt-padding} involves merely a constant factor
blowup. It can easily be generalized to arbitrary relational
structures. Applying Theorem~\ref{thm:alt-padding} to the CFI
graphs, we obtain graphs on which Duplicator wins the
$(\Alt{n},k)$-pebble game.

\begin{corollary}
\label{cor:alt-cfi}
There are families of graphs $(G_n)_n$ and $(H_n)_n$, interpreted
as $\taugraph$-structures, where $G_n$ and $H_n$ have
vertices $V$ with $\abs{V}=O(n)$ such that each $G_n$ and $H_n$ are
not isomorphic and Duplicator wins the
$(\Alt{V},k)$-pebble game on $G_n$ and $H_n$ for
some $k=\Omega(n)$. Furthermore, there is a polynomial-time
algorithm which outputs ``yes'' on every $G_n$ and ``no'' on every
$H_n$.
\end{corollary}

\begin{proof}
Apply the transformation of Theorem~\ref{thm:alt-padding} to the
CFI graphs of Theorems~\ref{thm:classical-cfi} and
\ref{thm:sym-cfi}. For the ``furthermore'' part, the algorithm
simply removes the $k+2$ isolated vertices and applies the algorithm
separating the CFI graphs of Theorem~\ref{thm:classical-cfi}.
\end{proof}

We call $(G_n)_n$ and $(H_n)_n$ the \defn{Alt-CFI graphs}.
There is a polynomial-time $\taugraph$-property,
the \defn{Alt-CFI property} $\PAltCFI$, such that
$G_n\in \PAltCFI$ and
$H_n\notin \PAltCFI$ for each $n$.

\subsection{The classical restriction of the partial trace game}

\begin{definition}
A \defn{base observable set} $\Obs$ for a group
$\Gamma\leq\Unit(\Hilb_n^\tau)$ is a set of observables on
$\Hilb_n^\tau$ which is closed under conjugation by $\Gamma$:
for every $O\in\Obs$ and every $U\in\Gamma$, $UOU^\dagger\in\Obs$.
\end{definition}

The classical base observable set for a classical unitary group
$\Gamma_G$ on a Hilbert space $\Hilb=(\C^2)^{\tprod m}$
of $m$ qubits is
$\ClasObs=\{O_i:i\in[m]\}$, where
\[
 O_i=\ket{1}_i\bra{1}_i\tprod \iden_{[m]\setminus\{i\}}.
\]
The subscripts indicate the qubits each tensor product component
acts upon. Each observable $O_i$ applies the projection
$\ketbra{1}{1}$ on the $i$th qubit and acts trivially on the
other qubits.

A set of pebbles $P\subseteq\Peb$ supports an observable
$O$ if every $U\in\inter P$ commutes with $O$. A unitary
$U\in\Gamma$ is an $\Obs$-partial isomorphism with respect
to $P$ from $\ket{A}$ to $\ket{B}$ if for every base observable
$O\in\Obs$ which is supported by $P$, we have
\[
 \braket[3]{A}{O}{A}=\braket[3]{B}{UOU^\dagger}{B}.
\]

\begin{definition}\label{def:observable-game}
The quantum $(\Gamma,\Peb,\Obs,k)$-pebble game on
$\ket{A}$ and $\ket{B}$ is defined as follows. The state of the game
consists of a subset of pebbles $P\subseteq\Peb$ with $\abs{P}\leq k$
and a unitary $U\in\Gamma$. Initially, no pebbles are placed and $U=\iden$.
Each round proceeds as follows.
\begin{enumerate}
\item Optionally, Spoiler removes a pebble from $P$.
If $\abs{P}=k$, Spoiler must remove a pebble.
\item Duplicator chooses a unitary $V\in\inter P$ and sets $U:=UV$.
\item Spoiler chooses a pebble from $\Peb\setminus P$ and adds it to $P$.
\end{enumerate}
Spoiler wins if $U$ is not an $\Obs$-partial isomorphism from
$\ket{A}$ to $\ket{B}$ with respect to $P$. Otherwise, play continues,
and Duplicator wins if it has a strategy to play forever.
\end{definition}

\begin{lemma}
\label{lem:classical-partial}
Let $\Gamma_G\leq\Unit(\Hilb_n^\tau)$ be a classical
unitary group, and let $\Peb_G$ be its classical pebble set.
Suppose $P\subseteq\Peb_G$ are the classical pebble groups
corresponding to the stabilizers of some set $P'\subseteq[n]$ of
domain elements. Let ${\str{A}},{\str{B}}$ be $\tau$-structures. If
$G=\Sym{n}$ or $\Alt{n}$ and
$\abs{P}\leq n-3$, then $U_\sigma\in\Gamma_G$ is an
$\ClasObs$-partial isomorphism with respect to
$P$ from $\ket{\str{A}}$ to $\ket{\str{B}}$ if and only if $\sigma$ is a
partial isomorphism with respect to $P'$ from ${\str{A}}$ to ${\str{B}}$.
\end{lemma}

\begin{proof}
Since $\abs{P'}=\abs{P}\leq n-3$ and $G=\Sym{n}$ or
$\Alt{n}$, every domain element $j\in[n]$ which is
not in $P'$ is moved by some even permutation
$\sigma\in\Stab_G(P')$, namely the cyclic
permutation $(j\ j'\ j'')$ between $j$ and two other distinct
domain elements $j',j''\notin P'$. So each base observable
$O_{(i,\vec{n})}\in\ClasObs$ is supported by
$P$ if and only if $\vec{n}\subseteq P'$. For each
$U_\sigma\in\Gamma_G$,
\[
 \braket[3]{\str{A}}{O_{(i,\vec{n})}}{\str{A}}
 =\braket[3]{\str{B}}{U_\sigma O_{(i,\vec{n})}U_\sigma^\dagger}{\str{B}}
\]
is equivalent to $\vec{n}\in R_i^{\str{A}}$ holding just when
$\sigma(\vec{n})\in R_i^{\str{B}}$, so the notions of partial isomorphism
are equivalent.
\end{proof}

\begin{lemma}
\label{lem:pt-classical}
Let $\Gamma_G$ be a classical unitary group, let
$P\subseteq\Peb_G$, and let
$\ket{\str{A}},\ket{\str{B}}\in\Hilb_n^\tau$ be computational basis
states encoding classical structures. $U\in\Gamma_G$ is an
$\ClasObs$-partial isomorphism from $\ket{\str{A}}$
to $\ket{\str{B}}$ if and only if it is a PT partial isomorphism from
$\ket{\str{A}}$ to $\ket{\str{B}}$, both with respect to $P$.
\end{lemma}

\begin{proof}
$P$ supports each
$O_{(i,\vec{n})}=\ket{1}_{(i,\vec{n})}\bra{1}_{(i,\vec{n})}\tprod \iden
\in\ClasObs$ if and only if qubit
$(i,\vec{n})$ is not moved by $\inter P$, that is,
$(i,\vec{n})\in Q_P^\tau$. Since $U\in\Gamma_G$ just permutes
the qubits,
\[
 \TrTo{Q_P^\tau}(\ketbra{\str{A}}{\str{A}})
 \quad\text{and}\quad
 \TrTo{Q_P^\tau}(U^\dagger\ketbra{\str{B}}{\str{B}} U)
\]
are both computational basis states, so they are equal if and only if
\[
 \braket[3]{\str{A}}{O_{(i,\vec{n})}}{\str{A}}
 =\braket[3]{\str{B}}{UO_{(i,\vec{n})}U^\dagger}{\str{B}}
\]
for all $(i,\vec{n})\in Q_P^\tau$, since $O_{(i,\vec{n})}$
measures qubit $(i,\vec{n})$ in the computational basis.
\end{proof}

\begin{corollary}
\label{cor:pt-equivalence}
On classical structures, the $(\Gamma_G,\Peb_G,k)$-PT
quantum pebble game is equivalent to the quantum
$(\Gamma_G,\Peb_G,\ClasObs,k)$-pebble
game.
\end{corollary}

We can then show that the PT quantum pebble game also cannot
distinguish the CFI graphs.

\begin{proposition}
\label{prop:cfi-pt}
Let $(G_n)_n$ and $(H_n)_n$ be the families of CFI graphs of
Theorem~\ref{thm:classical-cfi}, interpreted as
$\taugraph$-structures, and suppose $G_n$ and $H_n$
have the same set of vertices $V$. Let $G=\Sym{V}$.
Then for $n\geq3$, Duplicator wins the
$(\Gamma_G,\Peb_G,k)$-PT quantum pebble game on $G_n$ and
$H_n$ for some $k=\Omega(n)$.
\end{proposition}

\begin{proof}
By Theorem~\ref{thm:sym-cfi}, Duplicator wins the
$(\Sym{V},n)$-pebble game on $G_n$ and $H_n$. Let
$k=n-2$. Reducing the number of pebbles only makes a pebble game
easier for Duplicator, so Duplicator also wins the
$(\Sym{V},k)$-pebble game on $G_n$ and $H_n$.
Since $n\leq\abs{V}-1$ by Theorem~\ref{thm:classical-cfi}, we have
$k\leq\abs{V}-3$, so by Lemma~\ref{lem:classical-partial},
$\ClasObs$-partial isomorphism is equivalent to
classical partial isomorphism in this setting with at most $k$
pebbles. Thus Duplicator wins the quantum
$(\Gamma_G,\Peb_G,\ClasObs,k)$-pebble
game on $G_n$ and $H_n$. By Corollary~\ref{cor:pt-equivalence},
it also wins the $(\Gamma_G,\Peb_G,k)$-PT quantum pebble
game.
\end{proof}

Similarly, by the same proof and Corollary~\ref{cor:pt-equivalence}:

\begin{proposition}
\label{prop:alt-cfi-pt}
Let $(G_n)_n$ and $(H_n)_n$ be the families of Alt-CFI graphs of
Corollary~\ref{cor:alt-cfi}, interpreted as
$\taugraph$-structures, and suppose $G_n$ and $H_n$
have the same set of vertices $V$. Let $G=\Alt{V}$.
Then for $n\geq3$, Duplicator wins the
$(\Gamma_G,\Peb_G,k)$-PT quantum pebble game on $G_n$ and
$H_n$ for some $k=\Omega(n)$.
\end{proposition}

\subsection{Circuit size and entanglement width}

The following lemma connects bounds on orbit size to bounds on circuit size and entanglement width.
We obtain a tradeoff between these two parameters for a given lower bound on orbit size.

\begin{lemma}
\label{lem:wire-orbit}
Let $G\leq\Sym{n}$. Let $(C_n)_n$ be a family of
$\Gamma_G$-symmetric output quantum $\tau$-circuits over a
gateset whose gates act on at most $r$ wires, where $r$ is constant.
For $1 \le \ell \le \abs{W_n}$, where $W_n$ is the wire set of $C_n$, we have
\[ \maxorb_\ell(C_n) \le (\max\{r \cdot s(C_n) + 1,\ell_\tau(n)\})^\ell. \]
In particular, for every function $f : \Z_+ \to \Z_+$ with $f(n) = \omega(\log n)$, if $s(C_n) = 2^{o(f(n))}$, then $\maxorb_\ell(C_n) = 2^{o(\ell \cdot f(n))}$.
\end{lemma}

\begin{proof}
Let $W_n = X_n \sqcup A_n$ be the wires of $C_n$, where $X_n$ are the input wires and $A_n$ are the ancilla wires.
Let $E_n$ be the set of ancilla wires on which no gate acts, excluding the designated output wire, and set $R_n=A_n\setminus E_n$.
Every circuit automorphism preserves $E_n$, and every permutation of $E_n$ fixing the other wires is an internal automorphism.
In the $\In(C)$-orbit $[\vec w]$ of an ordered tuple $\vec w$, the positions occupied by wires in $E_n$ are thus fixed, while the particular unused wires in those positions can be permuted freely.
Hence we can remove those positions from $\vec w$ without changing the size of the orbit of $[\vec w]$ under $\Aut(C)/\In(C)$.

Hence, choose a tuple $\vec w$ such that the size of the $\Aut(C)/\In(C)$-orbit of $[\vec w]$ is maximized, and suppose $\vec w$ has $a$ positions containing wires in $R_n$ and $b$ positions containing input wires.
Then $a + b \le \ell$.
We have
\[
 \maxorb_\ell(C_n) \le \abs{R_n}^a\abs{X_n}^b \le \max\{\abs{R_n}, \abs{X_n}\}^\ell.
\]
For the vocabulary $\tau$, there are $\abs{X_n} = \ell_\tau(n)$ input wires.
At most $r\cdot s(C_n)$ nonempty ancilla wires are acted on by gates.
Retaining the output wire contributes at most one further wire, so
$\abs{R_n} \le r \cdot s(C_n) + 1$.
It follows that
\[
  \maxorb_\ell(C_n) \le \left(\max\{r \cdot s(C_n) + 1, \ell_\tau(n)\}\right)^\ell.
\]
In particular, if $s(C_n) = 2^{o(f(n))}$ for $f(n) = \omega(\log n)$, then $\maxorb_\ell(C_n) = 2^{o(\ell \cdot f(n))}$ since $\ell_\tau(n) = \poly(n) = 2^{o(f(n))}$ and $r$ is constant.
\end{proof}

For graph inputs and gates acting on at most two wires, we obtain the bound
\begin{equation}
 \maxorb_\ell(C_n) \le \left(\max\{2s(C_n) + 1, n^2\}\right)^\ell
 \label{eq:finite-orbit-bound}
\end{equation}
for $1 \le \ell \le \abs{W_n}$.
This identifies the circuit-size contribution to the lower-bound argument. The support theorem then relates these orbit sizes to the number of pebbles needed in the indistinguishability theorem.

\begin{theorem} \label{thm:general-lower-bound}
Let $G = \Alt{n}$.
Let $(k_n)_n$ and $(\ell_n)_n$ be sequences of positive integers such that $k_n \le n$ and $k_n/\ell_n = \omega(\log n)$.
Suppose that $(\str{A}_n)_n$ and $(\str{B}_n)_n$ are
families of $\tau$-structures of order $n$ such that for infinitely many $n$,
Duplicator wins the $(\Gamma_G,\Peb_G,k_n)$-$\ell_n$-PT
quantum pebble game on $\ket{{\str{A}}_n}$ and $\ket{{\str{B}}_n}$.
Let $\prop{P}$ be any property separating $(\str{A}_n)_n$ and $(\str{B}_n)_n$.
Then, there is no family of $\Gamma_G$-symmetric quantum $\tau$-circuits $(C_n)_n$ computing $\prop{P}$ which, for sufficiently large $n$, has both
entanglement width
\[ \ew(C_n) \le \ell_n \]
and size
\[ s(C_n) \le 2^{\min\{k_n/2, n/4-1\}/\ell_n - 1} - \frac{1}{2}. \]
\end{theorem}
\begin{proof}
Let $(C_n)_n$ be a family of such quantum circuits computing ${\prop{P}}$, and suppose $\ew(C_n) \le \ell_n$.
By the partition-symmetry result (Theorem~\ref{thm:partition-symmetry}), each $C_n$ has
$\ew(C_n)$-partition symmetry. The group
$\Alt{n}$ has distinct stabilizers for $n\geq4$.
Then by the indistinguishability theorem (Theorem~\ref{thm:indistinguishability}), for $n\geq4$, if
$\maxsupp_{\ew(C_n)}(C_n)\leq k_n/2$ and
Duplicator wins the
$(\Gamma_G,\Peb_G,k_n)$-$\ew(C_n)$-PT quantum pebble game
on $\ket{{\str{A}}_n}$ and $\ket{{\str{B}}_n}$, then
$C_n[\ket{{\str{A}}_n}]=C_n[\ket{{\str{B}}_n}]$, contradicting the assumption
that $(C_n)_n$ computes ${\prop{P}}$. Hence we must have
$\maxsupp_{\ew(C_n)}(C_n)>k_n/2$ for infinitely
many $n$.

By the quantum support theorem (Theorem~\ref{thm:support}), either $\maxorb_{\ew(C_n)}(C_n) > 2^{n/4-1}$ or $\maxsupp_{\ew(C_n)}(C_n) \le \log_2 \maxorb_{\ew(C_n)}(C_n)$.
It is clear that $\maxorb_\ell(C_n)$ is non-decreasing with $\ell$. %
Hence we obtain
\begin{equation}
    \maxorb_{\ell_n}(C_n) \ge \maxorb_{\ew(C_n)}(C_n) > 2^{\min\{k_n/2, n/4-1\}}
\end{equation}
for infinitely many $n$.
Applying Lemma~\ref{lem:wire-orbit} with $r = 2$, we get
\begin{equation} \label{eqn:general-lower-bound-intermediate}
    \max\{2s(C_n) + 1, \ell_\tau(n)\} > 2^{\min\{k_n/2,n/4-1\}/\ell_n}.
\end{equation}
Since $k_n/\ell_n = \omega(\log n)$, the right-hand side of (\ref{eqn:general-lower-bound-intermediate}) eventually exceeds $\ell_\tau(n) = \poly(n)$.
Hence we can drop $\ell_\tau(n)$ from the left-hand side, obtaining the bound
\[ s(C_n) > 2^{\min\{k_n/2,n/4-1\}/\ell_n-1} - \frac{1}{2} \]
which holds for infinitely many $n$.
\end{proof}

\begin{proof}[Proof of Theorem~\ref{thm:mainAlt}]
Write $m$ for the construction index of the Alt-CFI pairs in Corollary~\ref{cor:alt-cfi}, and let $N_m$ be their actual number of vertices. The original CFI graphs have between $m+1$ and $O(m)$ vertices, and the padding adds only $O(m)$ vertices. Thus $N_m=\Theta(m)$ and the orders $N_m$ are unbounded.

By Proposition~\ref{prop:alt-cfi-pt}, Duplicator wins the full
$(\Gamma_{\Alt{N_m}},\Peb_{\Alt{N_m}},k_{N_m})$-PT quantum pebble game on these pairs for $k_{N_m}=\Omega(m)=\Omega(N_m)$. It consequently wins the corresponding $\ell$-PT game for every $\ell\geq1$.

Apply Theorem~\ref{thm:general-lower-bound} to any family of $\Gamma_{\Alt{n}}$-symmetric output quantum circuits $(C_n)_n$ which separates the CFI pairs at each $N_m$.
For the conditions of that theorem, let $(\ell_n)_n$ be any sequence of integers with $\ell_n = O(n/f(n))$, and take $k_n = \Theta(n)$ where $k_{N_m}$ is as above.
Then $k_n/\ell_n = \Omega(f(n)) = \omega(\log n)$.
We obtain that there is no such family of circuits with both entanglement width $\ew(C_n) = O(n/f(n))$ and size $s(C_n) = 2^{o(f(n))}$.
\end{proof}

We remark on the constant factors in Theorem~\ref{thm:mainAlt}, Theorem~\ref{thm:main}, and its corollaries.
The quantitative lower bound on the size of symmetric circuits distinguishing the Alt-CFI graphs is determined by the constant $c$ such that for large enough $n$, Duplicator wins the $(\Alt{n},k)$-pebble game for some $k \ge cn$, where $n$ is the number of vertices. %
We give a lower bound on $c$.

A calculation by Pikhurko, Veith, and Verbitsky \cite[sec.~7.5]{Pikhurko2003} shows that on the original CFI graphs, for a large enough number of vertices $n'$, Duplicator wins the $(\Sym{n'},k)$-pebble game for some $k \ge 0.00465n'$.
In particular, Duplicator wins for $k = \ceil{0.00465n'}$ pebbles.
The construction in Corollary~\ref{cor:alt-cfi} increases the number of vertices by $k+2$:
hence using this value of $k$, the Alt-CFI graphs have $n \le 1.00465n' + 3$ vertices.
Hence, with this construction, Duplicator wins the $(\Alt{n},k)$-pebble game on the Alt-CFI graphs for $k \ge \frac{0.00465}{1.00465}(n-3) \ge 0.00462n$ pebbles for large enough $n$, so we obtain a lower bound of $c \ge 0.00462$.

Theorem~\ref{thm:general-lower-bound} then gives the quantitative lower bound
\[ s(C_n) > 2^{0.00231n/\ell_n-1} - \frac{1}{2} \]
on the size of any family of $\Gamma_{\Alt{n}}$-symmetric quantum circuits with entanglement width $\ew(C_n) \le \ell_n = o(n/\log n)$ distinguishing the Alt-CFI graphs; this inequality must be satisfied for infinitely many $n$.

This constant could be improved further, however, by using a more modern variant of the CFI construction: in particular, Schneider and Schweitzer \cite{Schneider2025} show the existence of pairs of graphs on which Duplicator wins the $(\Sym{n},k)$-pebble game for $k \ge 0.0105027n - o(n)$.
(Their argument, however, is nonconstructive.)
They study the equivalent notion of \defn{Weisfeiler--Leman dimension} (WL-dimension): this is the minimum $k$ such that the $k$-dimensional Weisfeiler--Leman test can distinguish a graph $G$ from any other non-isomorphic graph.
It is well-known that Duplicator wins the $(\Sym{n},k)$-pebble game on $G$ and $H$ if and only if $G$ and $H$ are not distinguished by the $(k-1)$-dimensional Weisfeiler-Leman test \cite{Cai1992}.
See e.g.~\cite{Kiefer2024} for more on lower bounds on the WL-dimension.

\section{Symmetric simulation of quantum algorithms}
\label{sec:symbqp}

We now show that, without restrictions on the entanglement width, it is possible to efficiently simulate every quantum algorithm that computes some graph property.
This result can be easily extended to general relational structures, but we focus on graphs for notational simplicity.

The first step is to show that threshold gates can be efficiently computed symmetrically.
For a set $S$ of $h$ control wires and an integer $t$, define $\mathrm{Thr}_{S,t}(x)=\mathbf{1}\{\sum_{u\in S}x_u\geq t\}$.
Its reversible implementation flips a separate target bit $z \notin S$ when $\mathrm{Thr}_{S,t}(x)=1$.
This function is clearly $\Sym{S}$-symmetric, and we wish to construct a $\Sym{S}$-symmetric quantum circuit that efficiently implements it.
In order to use our construction in the no-entanglement setting of Theorem~\ref{thm:ew-1-computes-fpc}, we additionally argue that our construction does not generate entanglement on computational-basis inputs.

\begin{lemma}[(Symmetric threshold computation)]
\label{lem:symbqp-threshold}
The map
\[
 \ket{x}\ket{z}
 \longmapsto
 \ket{x}\ket{z\oplus \mathrm{Thr}_{S,t}(x)}
\]
has a $\Sym{S}$-symmetric implementation using $O(h\log h)$ one-qubit and two-qubit gates and $O(\log h)$ workspace qubits.
Furthermore, if the input is in the computational basis, then the state after every gate in this construction is a product state.
\end{lemma}

\begin{proof}
Write $w = \sum_{u\in S}x_u$, so that $\mathrm{Thr}_{S,t}(x) = \mathbf{1}\{w\geq t\}$.
We first design a $\Sym{S}$-symmetric quantum circuit that writes the binary expansion of $w$ onto $m = \lceil \log_2 (h+1) \rceil$ designated ancilla wires.
From there, we can decide whether $w\geq t$ in an arbitrary way, as these ancilla wires are dissociated from the symmetries of $S$.

The circuit for obtaining the digits of $w$ uses only Hadamard gates $H$ and controlled phase gates $cR_j$, where
$R_j = \begin{pmatrix} 1 & 0\\ 0 & e^{2\pi i/2^j} \end{pmatrix}$.
First, let $c_1,\ldots,c_m$ be ancillary ``counter'' qubits starting at $\ket{0}$, and apply $H$ to each of them to create a uniform superposition.
Then, for every $u\in S$ and $1\leq j \leq m$, apply the controlled phase gate $cR_j$ with control qubit $u$ and target qubit $c_j$.
Note that these $hm$ two-qubit gates commute, even when they share a wire, so they form one layer.
This layer is $\Gamma_{\Sym{S}}$-symmetric, as any permutation of the qubits indexed by~$S$ only permutes the gates within this layer.

After these operations, the counter is now
\begin{equation}
 \label{eq:symbqp-fourier-counter}
 \bigotimes_{j=1}^{m} \frac{\ket{0}+ e^{2\pi iw/2^j}\ket{1}}{\sqrt2} = \frac{1}{\sqrt{2^m}} \sum_{k=0}^{2^m-1} e^{2\pi iwk/2^m} \ket{k}.
\end{equation}
This is precisely $F_{2^m} \ket{w}$, where $F_{2^m}$ is the quantum Fourier transform modulo $2^m$.
By applying its inverse $(F_{2^m})^\dagger$ to the counter qubits, we obtain $\ket{w}$.
Note that $\iden_S\otimes (F_{2^m})^\dagger$ is $\Gamma_{\Sym{S}}$-symmetric, as every unitary $U_\pi \otimes \iden$ for $\pi\in \Sym{S}$ is an automorphism of this circuit.

Finally, one can easily construct a reversible circuit that compares the two $m$-bit integers $w, t$ using $O(m^2)$ Toffoli and NOT gates, and $O(m)$ workspace bits.
Since a Toffoli gate can be implemented using five two-qubit gates, this leads to a quantum circuit for deciding whether $w\geq t$ which uses $O(m^2)$ two-qubit gates and $O(m)$ ancillas.
Again, as this circuit acts only on the counter ancillas, it is trivially $\Sym{S}$-symmetric.

Once this has been done, copy the result using a CNOT to the output wire and undo the whole computation to bring the non-output ancilla qubits back to $\ket{0}$.
The first part of the result follows.

To argue that the state is maintained in a product state when the input is in the computational basis, we analyze the standard implementations of $(F_{2^m})^\dagger$ and the Toffoli gate.
First, observe that the control wire of each $cR_j$ gate is in the computational basis when the input is, so no entanglement is produced in the first layer, even between the commuting gates.  %
We use the inverse quantum Fourier transform defined in \cite[chap.~5]{Nielsen2010}; each gate receives its own layer.
Observe that when the initial state is of the form in (\ref{eq:symbqp-fourier-counter}), the control qubits of each two-qubit gate in this construction are in the computational basis: namely, they are the bits of $w$ decoded so far.
Thus, no entanglement is created.

We use the five-gate Toffoli implementation of \cite[Fig.~4.8]{Nielsen2010}, where again each gate is in its own layer.
When the input is in the computational basis, so are the control qubits of each two-qubit gate, so no entanglement is created.
This yields the second part of the result.
\end{proof}

We can now prove Theorem~\ref{thm:symbqp-simulation}, which shows we can symmetrically simulate any quantum circuit on graph inputs with only a small overhead.

\begin{proof}[Proof of Theorem~\ref{thm:symbqp-simulation}]
By first copying the (computational-basis) input several times, running the algorithm on each copy with fresh workspace, and then taking a majority vote on the output qubits, we can assume that the quantum circuits to be simulated are correct with probability at least~0.9.

Write $r = \lceil\log_2(10n^2)\rceil$ and $M=2^r$.
For each vertex $v\in [n]$, add an $r$-qubit ``label'' register and apply a Hadamard to each label qubit.
These $rn$ gates form a single layer and prepare the state
\begin{equation} \label{eq:lambda}
    \frac{1}{\sqrt{M^n}} \sum_{\lambda: [n]\to \{0,\dots, M-1\}} \ket{\lambda}
\end{equation}
on the label registers.
We will use these labels to create an ordering of the vertices (in superposition) which does not break the symmetry of the circuit.

First, we describe a classical computation for a fixed label assignment.
For each ordered pair $(u,v)$ with $u\neq v$, compute $c_{u,v} = \mathbf{1}\{\lambda(u)<\lambda(v)\}$ using the same binary comparator for every pair, and set $c_{u,u}=0$.
A binary comparator uses $O(r)$ Boolean gates, and acts symmetrically with respect to the vertices.
For every $v\in [n]$ and $j\in [n]$, compute the rank indicator
$$h_{v,j} = \mathbf{1}\bigg\{\sum_{u\in [n]} c_{u,v} = j-1\bigg\}.$$
If all labels are distinct, then exactly one vertex has each rank $j$.
Denote the rank of $v$ by $\theta_\lambda(v)$, so $\theta_\lambda \in \Sym{n}$ is a permutation if $\lambda$ is injective.

For each pair $(i,j)\in [n]^2$, compute onto a fixed workspace wire
$$y_{i,j} = \bigvee_{u,v\in [n]} (x_{u,v} \wedge h_{u,i} \wedge h_{v,j}).$$
When $\lambda$ is injective, exactly one pair $(u,v)$ contributes to each $(i,j)$, and $y=b(\theta_\lambda \str{A})$.
In this case, these wires contain an isomorphic copy of the input, permuted by $\theta_\lambda$.

The entire classical computation above is symmetric:
a permutation $\sigma\in \Sym{n}$ sends a comparator indexed by $(u,v)$ to the one indexed by $(\sigma(u), \sigma(v))$, and sends $h_{v,j}$ to $h_{\sigma(v),j}$.
These rules also specify the action on internal comparator and Boolean clauses.
\cite[Proposition~1]{CastroSilva2026} shows how one can implement a symmetric classical circuit with threshold gates as a symmetric \emph{quantum} circuit with threshold gates without blowing up its size.
Using Lemma~\ref{lem:symbqp-threshold}, we can then implement it as a symmetric quantum circuit using only 1-qubit and 2-qubit gates.
Applying it to state~\eqref{eq:lambda}, we obtain
$$\frac{1}{\sqrt{M^n}} \sum_{\lambda: [n]\to \{0,\dots, M-1\}} \ket{\lambda} \ket{b(\theta_\lambda \str{A})}.$$

Run the given quantum algorithm on the second register using fresh workspace.
The resulting state can be written as
$$\frac{1}{\sqrt{M^n}} \sum_{\lambda: [n]\to \{0,\dots, M-1\}} \ket{\lambda} \ket{\Phi_{\lambda, \str{A}}},$$
where measuring the output qubit of $\ket{\Phi_{\lambda, \str{A}}}$ gives the correct result with probability at least~0.9 (by assumption).
Since this measurement acts trivially on the labels, orthogonality of the~$\ket{\lambda}$ gives the uniform average, so we obtain the correct result with probability~0.9 whenever the function $\lambda$ is injective.
Since the probability that a uniformly chosen $\lambda: [n] \to \{0, 1, \dots, M-1\}$ is \emph{not} injective is at most $\binom{n}{2}\frac{1}{M} \leq 1/20$, it follows from the union bound that our symmetric quantum algorithm is correct with probability at least $0.85 > 2/3$ on any input.
The theorem follows.
\end{proof}

Lemma~\ref{lem:symbqp-threshold} also gives a proof of Theorem~\ref{thm:ew-1-computes-fpc} by showing that symmetric quantum circuits of entanglement width~$1$ can simulate symmetric classical threshold circuits. A classical threshold circuit on graphs is a directed acyclic graph with an input gate for each potential edge, internal gates computing threshold or NOT functions, and a designated output gate. Such a circuit is symmetric if every permutation of the vertices of the input graph induces a permutation of the input gates that extends to an automorphism of the circuit, preserving the types and thresholds of the internal gates and fixing the output gate. We refer the reader to~\cite{CastroSilva2026} for the precise definitions.

\begin{lemma}[(No-entanglement simulation of classical threshold circuits)]
Let $\prop{P}$ be a graph property.
Suppose that a polynomial-time uniform family of symmetric classical circuits using threshold and NOT gates computes $\prop{P}$.
Then $\prop{P}$ can also be computed by a polynomial-time uniform family of symmetric quantum circuits with entanglement width 1 using only one-qubit and two-qubit gates.
\end{lemma}
\begin{proof}
Given a symmetric classical circuit, use Proposition 1 of \cite{CastroSilva2026} to convert it to a symmetric reversible circuit.
Each threshold gate can be implemented using Lemma~\ref{lem:symbqp-threshold} and each NOT gate by an $X$ gate.
\end{proof}

Combining this lemma with Anderson and Dawar's characterization of $\FPC$ (Theorem~\ref{thm:anderson-dawar-fpc}) gives Theorem~\ref{thm:ew-1-computes-fpc}.

\section{Acknowledgements}

Davi Castro-Silva thanks Anuj Dawar for insightful discussions on the computational power of symmetric quantum circuits.
Ryan Dancy is supported by a Clarendon Fund Scholarship from the University of Oxford.
Sergii Strelchuk acknowledges support from the Wellcome Leap as part of the Q4Bio Program and the Royal Society University Research Fellowship.

\appendix
\section{Symmetric implementation of QAOA}
\label{app:qaoa}

\begin{proof}[Proof of Theorem~\ref{thm:qaoa-compilation}]
Fix $m,n,p\geq1$, a threshold $t\in\{0,\ldots,m\}$, and the
angle vectors $\boldsymbol\gamma,\boldsymbol\beta\in\R^p$.
For an instance $J=(A,d)$ and an assignment $y\in\{0,1\}^n$, we define the cost function that corresponds to the number of satisfying equations
\begin{equation}
c_J(y)=\bigl|\{i\in[m]:(Ay)_i=d_i\}\bigr|.
\end{equation}
Following Farhi, Goldstone and Gutmann~\cite{Farhi2014},
we define the cost and mixing operations by $U_C(J;\gamma)\ket y=e^{-\mathrm{i}\gamma c_J(y)}\ket y$ and $U_M(\beta)=\prod_{j=1}^n e^{-\mathrm{i}\beta X_{y_j}}$ respectively,
where $X_i$ is the $X$ gate applied to the $i$-th qubit. Starting from
$\ket{\psi_0(J)}=\ket+^{\tprod n}$, the states after successive rounds are
\begin{equation}
\ket{\psi_\ell(J)}
=U_M(\beta_\ell)U_C(J;\gamma_\ell)\ket{\psi_{\ell-1}(J)},
\qquad \ell\in[p].
\end{equation}

Let $x_{ij}$ and $z_i$ be the input wires holding $A_{ij}$ and $d_i$.
We introduce assignment wires $y_j$, product wires $w_{ij}$, parity
wires $r_i$, and an output wire $w_{\out}$, each initialized in $\ket0$.
We call the product and parity wires, together with the counter
and comparison wires introduced below, the workspace.
All workspace wires are initialized in $\ket0$.
The input, assignment, and output wires are separate from
the workspace.
Apply a Hadamard gate to every $y_j$ in one layer to prepare $\ket+^{\tprod n}$.

Each Toffoli gate in the construction can be implemented using five
two-qubit gates. To see this, define
\begin{equation}
V=\frac{1+\mathrm{i}}2\iden+\frac{1-\mathrm{i}}2X,
\end{equation}
so that $V^2=X$. Let $C_b(V_u)$ denote the application of $V$ to target qubit $u$,
controlled by qubit $b$. The Toffoli gate with controls $a,b$
and target $u$ is
\begin{equation}
\operatorname{Toffoli}_{a,b\to u}
=
C_a(V_u)\,
\CNOT_{a\to b}\,
C_b(V_u^\dagger)\,
\CNOT_{a\to b}\,
C_b(V_u)
\end{equation}
This sequence restores the
control qubits and, for computational basis values $a,b\in\{0,1\}$,
applies
\begin{equation}
V^{\,b-(a\oplus b)+a}=V^{2ab}=X^{ab}
\end{equation}
to the target. By linearity, it therefore implements the Toffoli
gate on arbitrary superpositions.

For every $(i,j)\in[m]\times[n]$, compute
$w_{ij}=A_{ij}y_j$ using a Toffoli with controls $y_j,x_{ij}$.
We apply each step of the above decomposition simultaneously to all
pairs $(i,j)$. This gives five commuting layers: gates involving
$(x_{ij},w_{ij})$ act on disjoint pairs, while the other layers have
shared controls $y_j$ and distinct targets. Every input wire
$x_{ij}$ is restored after these five layers.

Next apply all $\CNOT_{w_{ij}\to r_i}$ in one commuting layer,
followed by all $\CNOT_{z_i\to r_i}$ in another. This computes
\begin{equation}
r_i=d_i\oplus\bigoplus_{j=1}^n A_{ij}y_j.
\end{equation}
Thus when equation $i$ is satisfied, we have $r_i=0$. Now apply
\begin{equation}
P_\gamma=\operatorname{diag}(e^{-\mathrm{i}\gamma},1)
\end{equation}
to each $r_i$ in one layer, and then reverse the parity and product
computations. For each computational basis state $\ket y$ of the assignment wires,
these operations multiply its amplitude by $e^{-\mathrm{i}\gamma c_J(y)}$ and restore all product wires $w_{ij}$ and parity wires $r_i$ to $\ket0$. This implements $U_C(J;\gamma)$ using $O(mn)$ gates in 15 commuting layers.

The mixer $U_M(\beta)$ is one layer of $n$ single-qubit gates.
Repeating these operations with angles $(\gamma_\ell,\beta_\ell)$
therefore leaves, after round $\ell$, the state
\begin{equation}
\ket J\ket{\psi_\ell(J)}
\ket0_{w_{\out}}\ket0_{\mathrm{workspace}}.
\end{equation}
The product and parity wires have returned to $\ket0$,
while the output, counter, and comparison wires have not
yet been used.
Preparation and all $p$ rounds use
$O(n+pmn)$ gates. Having prepared the QAOA state, we now compute whether the assignment satisfies at least $t$ equations and record the result on the output wire. If $t=0$, we apply $X$ to the output wire. For $1\leq t\leq m$, we first compute the number of satisfied equations.

\paragraph{Computing the number of satisfied equations.}
Set $1\leq t\leq m$. We now check which equations are satisfied by the assignment. The product and parity wires were restored to $\ket{0}$ after each QAOA round, so we can reuse them. For each $i,j$, compute $A_{ij}y_j$ on $w_{ij}$. Then compute
\begin{equation}
r_i=d_i\oplus\bigoplus_{j=1}^n A_{ij}y_j.
\end{equation}
The value of $r_i$ is $0$ when the two sides of the $i$th
equation agree, and $1$ when they differ. Applying $X$ to
$r_i$ reverses these values. Thus this wire now holds
\begin{equation}
s_i=
\begin{cases}
1 & \text{if the $i$th equation is satisfied by $y$,}\\
0 & \text{otherwise.}
\end{cases}
\end{equation}
The number of satisfied equations is $c_J(y)=\sum_{i=1}^m s_i$.
Computing all the values $s_i$ uses $O(mn)$ gates.

We compute the sum by first recording it in the phases of a
counter register, then converting those phases into binary digits.
This allows all equations to contribute to the counter in a single
commuting layer, without choosing an order for them. The counter must represent the integers from $0$ to $m$, so we use
$L=\lceil\log_2(m+1)\rceil$ wires and set $K=2^L$.
Introduce counter wires $q_0,\ldots,q_{L-1}$.
Apply a Hadamard gate to each wire to change its initial
state $\ket0$ to $\ket+$.
For every $i\in[m]$ and $0\leq j<L$, apply
\begin{equation}
D_j=\operatorname{diag}
\bigl(1,1,1,e^{2\pi\mathrm{i}2^j/K}\bigr)
\end{equation}
to $(r_i,q_j)$. This gate does nothing when $r_i$ holds $0$.
When $r_i$ holds $1$, it multiplies the $\ket1$ component of
$q_j$ by $e^{2\pi\mathrm{i}2^j/K}$. Thus each satisfied equation
makes the same contribution to the phase of $q_j$.
All these gates are diagonal, so they commute and form a single
layer.

For an assignment $y$ satisfying $k=c_J(y)$ equations, these
contributions multiply to give
\begin{equation}
\ket{\chi_k}
=\bigotimes_{j=0}^{L-1}
\frac{\ket0+e^{2\pi\mathrm{i}2^j k/K}\ket1}{\sqrt2}.
\end{equation}
The number $k$ is now encoded in the relative phases of the
counter qubits. We next convert it into binary, writing
\begin{equation}
k=\sum_{a=0}^{L-1}k_a2^a,
\end{equation}
where each $k_a\in\{0,1\}$.

The least significant digit is particularly easy to obtain.
Since $K=2^L$, the state of $q_{L-1}$ is
\begin{equation}
\frac{\ket0+e^{\pi\mathrm{i}k}\ket1}{\sqrt2}
=
\frac{\ket0+(-1)^{k_0}\ket1}{\sqrt2}.
\end{equation}
A Hadamard gate maps this state to $\ket{k_0}$.
We can then use this qubit as a control to remove the contribution
of $k_0$ from the phases of the remaining counter qubits.
Repeating this procedure recovers one binary digit at each step.

More precisely, at step $a=0,\ldots,L-1$, apply a Hadamard gate
to $q_{L-1-a}$. Then apply
\begin{equation}
\operatorname{diag}
\bigl(1,1,1,e^{-2\pi\mathrm{i}2^{j+a}/K}\bigr)
\end{equation}
to each pair $(q_{L-1-a},q_j)$ with $0\leq j<L-1-a$.
Once $q_{L-1-a}$ holds $k_a$, these gates multiply the $\ket1$
component of each remaining $q_j$ by
$e^{-2\pi\mathrm{i}2^j k_a2^a/K}$, cancelling the contribution
of the term $k_a2^a$.

To verify the procedure by induction, suppose that the
contributions of $k_0,\ldots,k_{a-1}$ have already been removed.
The relative phase on $q_{L-1-a}$ is then
\begin{equation}
\exp\left(
\frac{2\pi\mathrm{i}2^{L-1-a}}{K}
\left(k-\sum_{b<a}k_b2^b\right)
\right)=(-1)^{k_a}.
\end{equation}
Indeed, the expression in parentheses is $k_a2^a$ plus a
multiple of $2^{a+1}$. The Hadamard gate therefore maps this
qubit to $\ket{k_a}$, and the subsequent phase gates remove
this digit's contribution from the remaining qubits.

At the end, $q_{L-1-a}$ holds $k_a$ for every $a$. The counter
therefore stores the number of satisfied equations in binary,
with the least significant bit on $q_{L-1}$. Recording the sum
in the phases uses $mL$ gates, and converting it to binary uses
$O(L^2)$ gates, all acting on the counter wires.
No measurements are needed: by linearity, the construction
also works on a superposition of assignments, associating each
assignment $\ket y$ with the binary representation of $c_J(y)$.

\paragraph{Comparing with the threshold.}
The counter now stores $k=c_J(y)$ in binary. We now compute whether $k\geq t$ and record the answer on the output wire, without changing the counter. Since $t\geq1$, we can test the equivalent condition $k>t-1$. We write
\begin{equation}
t-1=\sum_{a=0}^{L-1}b_a2^a,
\end{equation}
where $b_a\in\{0,1\}$ are fixed by the threshold. To compare two binary integers, consider the most significant position at which their digits differ. The larger integer has $1$ at that position, and the smaller has $0$. Thus $k>t-1$ precisely when there is a position $j$ at which $k_j=1$ and $b_j=0$, with all more significant digits equal. For each position $j$ with $b_j=0$, we therefore compute
\begin{equation}
e_j(k)=k_j\wedge
\bigwedge_{a=j+1}^{L-1}(k_a=b_a).
\end{equation}
It equals $1$ when $j$ is the most significant position at which the two integers differ and $k$ has the larger digit. When $j=L-1$, there are no more significant digits to check, so $e_j(k)=k_j$.

At most one of these expressions can be $1$, since two integers have at most one most significant position at which they differ. When $k\geq t$, one of these expressions equals $1$ and all the others are $0$. When $k<t$, they all equal $0$, thus
\begin{equation}
\bigoplus_{j:b_j=0}e_j(k)
=\mathbf{1}_{\{k\geq t\}}.
\end{equation}

Introduce auxiliary wires for the comparison, which we call
the comparison wires. These start in $\ket0$ and are reused
to compute the expressions $e_j(k)$ one at a time.

Each expression is a conjunction of at most $L$ conditions
on the counter bits. A condition $k_a=b_a$ asks for $k_a=1$
when $b_a=1$, and for $k_a=0$ when $b_a=0$.
In the latter case, we temporarily apply $X$ to the
corresponding counter wire, so that every condition can
be tested by a control on $1$.

A sequence of Toffoli gates computes the conjunction on a comparison wire, using other comparison wires for intermediate results. For a conjunction consisting of a single bit, a CNOT suffices. We use the wire holding $e_j(k)$ to control a CNOT onto $w_{\out}$, then apply the gates that computed the conjunction in reverse order and undo the $X$ gates on
the counter wires. This restores the counter and returns all comparison wires to $\ket0$, ready for the next value of $j$.

Each expression requires $O(L)$ gates, since every Toffoli uses the five two-qubit gates described above. There are at most $L$ expressions, so the comparison uses $O(L^2)$ gates.
The output wire starts in $\ket0$ and is flipped once for each expression whose value is $1$. It therefore records their exclusive OR, which is $1$ precisely when $k\geq t$.

Having recorded the comparison result, we no longer need the counter or the values indicating which equations are satisfied. We undo the operations that computed them, leaving
the output wire untouched. All workspace wires return to $\ket0$.

Recall the QAOA state has the form
\begin{equation}
\ket{\psi_p(J)}=\sum_{y\in\{0,1\}^n}a_y\ket y,
\end{equation}
and the final state can be succinctly described as
\begin{equation}
\ket J
\sum_{y\in\{0,1\}^n}
a_y\ket y
\ket{\mathbf{1}_{\{c_J(y)\geq t\}}}_{w_{\out}}
\ket0_{\mathrm{workspace}}.
\end{equation}
Here the indicator is $1$ when $c_J(y)\geq t$ and $0$
otherwise.
Measuring the output wire gives a required acceptance probability
\begin{equation}
C[\ket J]
=\sum_{y:c_J(y)\geq t}|a_y|^2.
\end{equation}

\paragraph{Symmetry.}
We now verify that relabelling the equations and variables preserves every layer of the circuit. Let $(\alpha,\sigma)\in\Sym{m}\times\Sym{n}$, where $\alpha$ permutes the equations and $\sigma$ permutes the variables. We extend this relabelling to the assignment, product, and parity wires:
\begin{equation}
\begin{aligned}
x_{ij}&\mapsto x_{\alpha(i),\sigma(j)},&
z_i&\mapsto z_{\alpha(i)},\\
w_{ij}&\mapsto w_{\alpha(i),\sigma(j)},&
r_i&\mapsto r_{\alpha(i)},\\
y_j&\mapsto y_{\sigma(j)}.
\end{aligned}
\end{equation}
Relabelling the equations and variables does not change the number of satisfied equations or the result of comparing this number with $t$. We extend the wire permutation by mapping each counter wire, each comparison wire, and the output wire $w_{\out}$ to itself. We now check that this choice preserves every circuit layer by considering its successive layers. Each of the five layers used to compute the products $A_{ij}y_j$ applies the same gate for every pair $(i,j)$. Relabelling these pairs permutes the gates within each layer. The parity computation, cost phases, and $X$ gates that mark satisfied equations treat all equations alike. Similarly, the Hadamard preparation and mixing operations apply the same gate to every assignment wire. Each of these layers is therefore preserved.

In the counting layer, each counter wire $q_j$ interacts with every equation wire $r_i$ through the same gate $D_j$. Permuting the equations only permutes these gates within the layer. The subsequent conversion to binary and comparison with $t$ act solely on counter, comparison, and output wires, all of which are fixed. Each gate in these computations is
therefore fixed, and we may place these gates in separate layers.

Finally, undoing a computation replaces its layers by their inverses in reverse order. Since the wire permutation preserves each original layer, it also preserves each inverse layer. For $t=0$, the final $X$ gate acts on the fixed output wire and is also preserved. Thus every independent permutation of the equations and variables extends to a permutation of all
circuit wires that preserves every layer and fixes the output.

\paragraph{Gate count.}
Preparing the assignment register uses $n$ Hadamard gates.
Each QAOA round uses $O(mn)$ gates for the cost operation and
$n$ gates for the mixer, giving $O(n+pmn)$ gates in total.

For the final threshold test, computing which equations are
satisfied uses $O(mn)$ gates. Recording their number in the
counter phases uses $mL$ gates, and converting the counter
to binary and comparing it with $t$ each use $O(L^2)$ gates.
Undoing the workspace computations adds at most the same
order of cost. The threshold test therefore uses
$O(mn+mL+L^2)$ gates. When $t=0$, only the output $X$ gate
is needed.

For $m\geq2$, we have $L=O(\log m)$ and
$L^2=O(m\log m)$. Since $p\geq1$, the total gate count is
\begin{equation}
s(C)=O\bigl(n+pmn+mn+mL+L^2\bigr)
     =O\bigl(pmn+m\log m\bigr).
\end{equation}
For $m=1$, we have $L=1$, so the counter and comparison
require only a constant number of gates. Their cost is
absorbed by $O(pmn)$, giving the same bound.
Every gate in the construction acts on at most two qubits.
\end{proof}
\section{Additional lemmas}

\begin{lemma} \label{lem:app-pure-induces-partition}
  Let $\Hilb = \Hilb_A \tprod \Hilb_B$ be a bipartite Hilbert space.
  Let $\rho \in \Dens(\Hilb)$.
  If $\Tr_B(\rho) = \ketbra\psi\psi$ is a pure state, then $\rho = \ketbra\psi\psi \tprod \rho'$, where $\rho' = \Tr_A(\rho)$.
\end{lemma}
\begin{proof}
  Let $\{\ket{a_i}\}_i$ and $\{\ket{b_i}\}_i$ be orthonormal bases for $\Hilb_A$ and $\Hilb_B$ respectively, chosen such that $\ket\psi = \ket{a_1}$.
  Write $\rho = \sum_i p_i \ketbra{\psi_i}{\psi_i}$ for some positive real numbers $\{p_i\}_i$ with $\sum_i p_i = 1$, and further write each $\ket{\psi_i}$ as $\ket{\psi_i} = \sum_{jk} \alpha_{ijk} \ket{a_j} \ket{b_k}$ for some complex numbers $\{\alpha_{ijk}\}_{jk}$ where $\sum_{jk} \abs{\alpha_{ijk}}^2 = 1$.
  We obtain
  \begin{align}
    \rho &= \sum_i p_i \left( \sum_{jk} \alpha_{ijk} \ket{a_j} \ket{b_k} \right) \left( \sum_{jk} \alpha_{ijk}^\ast \bra{a_j} \bra{b_k} \right) \\
    &= \sum_{i,j,k,j',k'} p_i \alpha_{ijk} \alpha_{ij'k'}^\ast \ketbra{a_j}{a_{j'}} \tprod \ketbra{b_k}{b_{k'}}
  \end{align}
  where $\alpha^\ast$ denotes the complex conjugate of $\alpha$.
  Hence,
  \begin{equation}
    \Tr_B(\rho) = \ketbra\psi\psi = \sum_{i,j,j',k} p_i \alpha_{ijk} \alpha_{ij'k}^\ast \ketbra{a_j}{a_{j'}}.
  \end{equation}
  Since we chose $\ket\psi = \ket{a_1}$, we therefore have
  \begin{equation}
    \sum_{ik} p_i \alpha_{i1k} \alpha_{i1k}^\ast = \sum_{ik} p_i \abs{\alpha_{i1k}}^2 = 1
  \end{equation}
  and since $\sum_i p_i = 1$, we must have $\sum_k \abs{\alpha_{i1k}}^2 = 1$ for each $i$.
  Since $\sum_{jk} \abs{\alpha_{ijk}}^2 = 1$ for each $i$, we therefore have $\alpha_{ijk} = 0$ whenever $j \ne 1$.
  Hence,
  \begin{align}
    \rho &= \sum_{i,k,k'} p_i \alpha_{i1k} \alpha_{i1k'}^\ast \ketbra\psi\psi \tprod \ketbra{b_k}{b_{k'}} = \ketbra\psi\psi \tprod \sum_{i,k,k'} p_i \alpha_{i1k} \alpha_{i1k'}^\ast \ketbra{b_k}{b_{k'}}
  \end{align}
  and so $\rho = \ketbra\psi\psi \tprod \Tr_A(\rho)$.
\end{proof}

\begin{lemma} \label{lem:app-pure-product-implies-pure}
  Let $\Hilb = \Hilb_A \tprod \Hilb_B$ be a bipartite Hilbert space.
  Let $\rho_1 \in \Dens(\Hilb_A)$ and $\rho_2 \in \Dens(\Hilb_B)$.
  Then $\rho_1 \tprod \rho_2$ is a pure state if and only if both $\rho_1$ and $\rho_2$ are pure states.
\end{lemma}
\begin{proof}
  For any density matrix $\rho$ we have $\Tr(\rho^2) \le 1$, with equality if and only if $\rho$ is a pure state \cite{Nielsen2010}.
  We have $\Tr((\rho_1 \tprod \rho_2)^2) = \Tr(\rho_1^2 \tprod \rho_2^2) = \Tr(\rho_1^2)\Tr(\rho_2^2)$, so $\Tr((\rho_1 \tprod \rho_2)^2) = 1$ iff $\Tr(\rho_1^2) = \Tr(\rho_2^2) = 1$.
\end{proof}

\begin{lemma} \label{lem:app-pure-overlap}
  Let $\Hilb = \Hilb_A \tprod \Hilb_B \tprod \Hilb_C$ be a tripartite Hilbert space.
  Let $\rho^{ABC} \in \Dens(\Hilb)$.
  If $\rho^{AB} = \Tr_{C}(\rho^{ABC})$ and $\rho^{AC} = \Tr_{B}(\rho^{ABC})$ are pure states, then $\rho^A = \Tr_{BC}(\rho^{ABC})$ is a pure state.
\end{lemma}
\begin{proof}
  Let $\ket\psi \in \Hilb \tprod \Hilb_D$ be a purification of $\rho^{ABC}$, \ie{} a state $\ket\psi$ such that $\Tr_{D}(\ketbra\psi\psi) = \rho^{ABC}$ for some Hilbert space $\Hilb_D$ \cite{Nielsen2010}.
  Since $\rho^{AB}$ is a pure state, by Lemma~\ref{lem:app-pure-induces-partition},
  \begin{equation}
    \ketbra\psi\psi = \rho^{AB} \tprod \Tr_{AB}(\ketbra\psi\psi).
  \end{equation}
  Furthermore, $\rho^{AC} = \Tr_{BD}(\ketbra\psi\psi)$, so
  \begin{equation}
    \rho^{AC} = \Tr_B(\rho^{AB}) \tprod \Tr_D(\Tr_{AB}(\ketbra\psi\psi)) = \rho^A \tprod \Tr_{ABD}(\ketbra\psi\psi).
  \end{equation}
  By Lemma~\ref{lem:app-pure-product-implies-pure}, since $\rho^{AC}$ is a pure state, $\rho^A$ is a pure state.
\end{proof}

\clearpage
\printbibliography[title={References}]
\end{document}